\documentclass[sigplan,nonacm]{acmart}
\usepackage{graphicx}
\usepackage{listings}
\usepackage{booktabs}
\usepackage{caption}
\usepackage{subcaption}
\usepackage{seqsplit}
\usepackage{url}
\usepackage{amsmath}
\usepackage{fancyhdr}
\usepackage{verbatim}
\usepackage{hyperref}
\usepackage{textcomp}
\usepackage{xcolor}

\usepackage{tabularx}
\usepackage{makecell}
\usepackage{adjustbox}
\usepackage{bbding}
\usepackage{pifont}
\usepackage{wasysym}
\usepackage{multirow}
\usepackage{siunitx}
\usepackage{csquotes}
\usepackage{enumitem}
\usepackage{comment}
\usepackage{nicefrac}
\usepackage[normalem]{ulem}

\newcommand{\para}[1]{\textbf{\textit{#1}}}
\newcommand{\oursys}{{\textsc{TileSight}}}
\newcommand{\secref}[1]{Sec.~\ref{#1}}

\usepackage{tikz}
\usepackage[linesnumbered,ruled,vlined,noend]{algorithm2e}

\begin{document}

\hyphenpenalty=600

\lstset{
  language=Python,
  basicstyle=\fontsize{7.5}{8.5}\ttfamily,
  keywordstyle=\color{blue},
  commentstyle=\itshape\color{green!40!black},
  frame=single,
  numbers=left,
  stepnumber=1,
  emph={getGridDim, getBlockDim, kernel, getBlockIdx, StreamModule, virtual, Schedule, Signal, GetBlockExecutorId},
  emphstyle=\textbf,
  morekeywords={class, void, size_t, vector, Map, string, sEUType,CacheConfig, struct, enum, ScratchConfig,std,function,TileType, TileShape, Tile, MemoryHierarchy, ComputeUnit},
  deletekeywords={bool},
  emph={[2] StreamModule, virtual, Schedule, Signal, GetBlockExecutorId},
  emphstyle={[2]\color{purple!80!black}}
}

\title[]{
\oursys{}: A First-Principles Tile-Centric Analytical GPU Performance Model from Cores to Clusters
}

\newcommand\paperauthors{
    Zhiwen Mo\textsuperscript{1},
    Yu Cheng\textsuperscript{2},
    Lei Wang\textsuperscript{2},
    Zhengju Tang\textsuperscript{2},
    Lei Xu\textsuperscript{3},
    Guoyu Li\textsuperscript{1},
    Yuqi Dong\textsuperscript{2},
    Lingxiao Ma\textsuperscript{4},
    Yuqing Xia\textsuperscript{4},
    Jilong Xue\textsuperscript{4},
    Fan Yang\textsuperscript{5},
    Luo Mai\textsuperscript{6},
    Zhi Yang\textsuperscript{2},
    Wayne Luk\textsuperscript{1},
    Hongxiang Fan\textsuperscript{1}
}

\newcommand{\paperaffiliation}{%
  \mbox{%
    \small
    \textsuperscript{1}Imperial College London,
    \textsuperscript{2}Peking University,
    \textsuperscript{3}Shanghai Jiao Tong University,
    \textsuperscript{4}Tile-AI,
    \textsuperscript{5}Microsoft Research,
    \textsuperscript{6}University of Edinburgh%
  }%
}

\author{\paperauthors}
\affiliation{\institution{\paperaffiliation}\country{}}

\renewcommand{\shortauthors}{Zhiwen Mo et al.}

\begin{abstract}
Recent GPU programming frameworks, such as Triton, TileLang, and CUDA Tile, have adopted the tile as a first-class language primitive, making tile-centric programming the prevailing approach for writing high-performance GPU kernels.
Performance-analysis tooling for tile-based programs, however, has not followed suit: programmers still fall back on coarse roofline bounds, opaque ML-based predictors, or post-hoc profilers to reason about how their kernels actually run.
This gap is increasingly painful for modern AI workloads, in which kernel fusion and distributed inference hinge on the interplay of tensor cores, CUDA cores, the cache hierarchy, memory pipelines, and inter-GPU networks.
We bridge this gap with \oursys{}, a tile-centric performance-modeling tool that leverages the tile from a programming primitive to an analysis primitive.
Within a single GPU core, \oursys{} models the overlap of compute and memory pipelines.
Across cores on the same chip, \oursys{} models the cache hierarchy. 
Across GPUs, it models inter-node communication.
All three layers share the tile abstraction:
\textit{1)} the intra-tile layer expresses each tile's work as a resource vector spanning the network, memory, and compute pipelines;
\textit{2)} the inter-tile layer schedules dependent and ordered tile actions to expose legal overlap and infers multi-level cache hit rates from a tile reuse distance;
\textit{3)} the cross-device layer maps remote tensor accesses to placements and routes them through an $\alpha$--$\beta$ stage cost.
Evaluated on A100, H200, B200, and B6000, \oursys{} matches measured single-GPU kernel latency to within a pooled 12.35\% mean absolute percentage error (MAPE), beating state-of-the-art baselines and transferring better across the four architectures.
Its L2 cache-hit-rate predictions land within roughly one percentage point of the measured rate on every GPU.
Pushed up to 32-GPU deployments, \oursys{} reaches 16.18\% weighted MAPE (wMAPE) on fused distributed kernels and 13.52\% wMAPE on end-to-end vLLM serving.
When driven into the optimization loop, \oursys{} picks tile configurations competitive with strong vendor and expert baselines on the case studies we report.
\oursys{} will be open-sourced upon publication.
\end{abstract}

\maketitle

\section{Introduction}
\label{sec:intro}

Large language model (LLM) scaling keeps pushing training and serving systems closer to hardware limits, making kernel efficiency central to both latency and cost.
To extract this performance, developers increasingly fuse multiple LLM operations into large tile programs whose bottlenecks are determined by tiling, memory movement, pipeline overlap, and wavefront scheduling.
Accurate and fast white-box performance modeling is therefore needed to expose the performance boundary and guide optimization.

To facilitate kernel optimization, the GPU programming community has converged on \emph{tile-centric programming} as the common paradigm: Triton~\cite{tillet2019triton} pioneered tile-level loads, stores and dot products, which has become the de facto standard for custom kernels in PyTorch. TileLang~\cite{tilelang} further decouples dataflow from scheduling at the tile level. NVIDIA's CUDA Tile~\cite{cudatile} (CUDA 13.1, 2025), described as the most significant CUDA advancement in roughly 20 years~\cite{cudatile_keynote}, officially adopts tiles as the programming primitive, while CuteDSL~\cite{cutlass} exposes CUTLASS's tile abstractions as a Python domain-specific language (DSL).
As a result, tiles have emerged as a central abstraction in modern GPU programming. However, \textbf{the performance analysis has not kept pace with this tile-centric abstraction}: Triton relies on black-box autotuning over thousands of configurations~\cite{tillet2019triton}, roofline models~\cite{williams2009roofline} cannot distinguish an L2 cache miss from a shared memory bank conflict, and ML-based predictors~\cite{lee2025forecasting,geoffrey2021habitat} require per-architecture training and are opaque.
Profilers such as Nsight Compute (NCU) and profiling-based tools~\cite{guan2025kperfir,huang2025neutrino} are post-hoc: they can perturb execution through instrumentation and clock changes, and their counter reports do not explain which tile, reuse pattern, or pipeline stage caused the observed bottleneck.
Table~\ref{tab:method_comparison} summarizes this abstraction mismatch.
As tile-centric programming becomes increasingly adopted, an accurate and efficient \emph{tile-centric performance model} is urgently needed to predict how tile-configuration changes affect performance without running the kernel.

The need for such a model becomes even more pressing when we consider what happens \emph{inside} a tiled kernel's main loop.
Even for GEMM, performance depends on software-pipeline depth, resident tiles per streaming multiprocessor (SM), and load-compute overlap rather than only on FLOP and byte counts.
The issue becomes sharper in fused kernels: as illustrated in Figure~\ref{fig:fa3_mapping}, FlashAttention-3 (FA-3) on H100 involves over ten distinct operations, including two GEMMs on Tensor Cores, reductions and softmax on CUDA cores, and special functions on special function units (SFUs), with fine-grained data dependencies.
These operations occupy different hardware resources and can potentially overlap, but the degree of overlap depends critically on their scheduling order and pipeline depth.
Existing tools, including roofline, profilers, and autotuners, are largely blind to this intra-tile scheduling structure. Furthermore, these complex kernels are increasingly needed in \emph{distributed} settings. For instance, Tensor parallelism (TP), expert parallelism (EP), and sequence parallelism (SP) partition workloads across multiple GPUs~\cite{svedas2025survey}, introducing collective communication that needs to overlap with computation. 
The performance of a distributed kernel depends on how the global tile grid is partitioned, what communication primitives are used, and how compute and communication pipelines interleave. These decisions are currently made by intuition or expensive trial-and-error.

To address these challenges, we observe that tiles provide \emph{a natural first-class abstraction for performance modeling} of GPU systems. This stems from three properties:
\textbf{(1) Deterministic}: given a tile configuration (shape, pipeline depth, memory layout), the resource usage of each tile is fully determined, enabling analytical modeling without simulation.
\textbf{(2) Composable}: tile information composes hierarchically. Each tile carries its own per-pipeline resource decomposition (intra-tile), tiles are related through dependencies, concurrent issue, and execution order (inter-tile), and tile grids extend across devices through placement (cross-device). Each level can be modeled independently then composed.
\textbf{(3) Portable}: the tile abstraction is adopted across various GPU architectures (in this paper we exercise NVIDIA A100, H100, H200, B200, RTX PRO 6000 Blackwell (B6000), and AMD MI210), since all modern GPUs execute tile-shaped workloads through similar hierarchical memory and compute structures.

Building on these insights, we present \oursys{}, a \emph{unified tile-centric analytical execution engine}. Unlike roofline models that attribute performance to a single bottleneck resource, \oursys{} analytically simulates how a tile execution plan unfolds across the hardware, capturing the prologue, steady-state overlap, and epilogue structure that determines real kernel performance. This simulation composes three hierarchical levels with unified tile-based abstractions:

\begin{itemize}[leftmargin=*, itemsep=1pt]
\item \textbf{Intra-tile}: each tile is characterized by an operation, an src/dst placement descriptor, and a footprint, which together produce a per-tile \emph{resource vector} that decomposes work into times on independently schedulable hardware pipelines spanning network, memory, and compute. The same placement descriptor unifies fusion (intermediates kept in registers or shared memory (SMEM)) and cross-device movement.

\item \textbf{Inter-tile}: tiles are related through producer--consumer dependencies, concurrent issue, and execution order. These together drive a topological-order search over the tile-action directed acyclic graph (DAG) that picks the best legal pipeline overlap inside a fused kernel body, and a multi-level \emph{tile reuse distance} analysis with stochastic distance-based cache modeling (SDCM) that derives implicit cache hit rates from grid traversal.

\item \textbf{Cross-device}: cross-device execution is a placement case of the same intra-tile abstraction --- a tile whose source or destination crosses devices simply gains a \texttt{Net} entry computed from the routed $\alpha$--$\beta$ cost of the underlying remote tensor access, so the same envelope still applies.
\end{itemize}

Crucially, these three levels are jointly designed with shared core abstractions: \texttt{HardwareUsage} as a per-pipeline time decomposition, the \emph{tile action} as the composable scheduling unit, and the \texttt{TileGrid} as the workload descriptor.
In summary, we make the following contributions:

\noindent
\textbf{(1) A unified tile-centric analytical execution engine} that simulates how tile execution plans unfold across hardware pipelines, including per-tile resource decomposition (intra-tile), dependency-driven DAG ordering with tile reuse-distance cache modeling (inter-tile), and placement-based cross-device tile accesses, all under one framework with shared abstractions (\S\ref{sec:design}).

\noindent
\textbf{(2) Tile-pipeline overlap analysis} that models both regular software pipelined loops (e.g., GEMM load-compute overlap) and complex fused kernels (e.g., FlashAttention and multi-head latent attention (MLA) decode) as repeated tile pipelines over dependency-constrained tile-action DAGs. By combining pipeline depth, resident tile interleaving, and legal tile-action ordering, \oursys{} predicts the prologue, steady-state, and epilogue costs that simple roofline models miss (\S\ref{ssec:pse}).

\noindent
\textbf{(3) Tile reuse-distance cache modeling} that makes cache behavior a natural consequence of the tile execution plan rather than a separate trace-simulation problem.
By reasoning about reuse at the same granularity as GPU schedules, \oursys{} enables fast, schedule-sensitive multi-level cache modeling inside an analytical performance model, while preserving accuracy through lightweight approximation and sampling techniques (\S\ref{ssec:cache_model}).

\noindent
\textbf{(4) Composable distributed extension via tile placement} where cross-device execution is a placement case of the same tile abstraction: remote tensor accesses are inferred from producer--consumer placement and decomposed into ordered stages of logical exchanges, whose routed $\alpha$--$\beta$ cost populates the network entry of the per-tile resource vector so cross-device movement composes with local compute through the same envelope (\S\ref{ssec:distributed}).

\section{Background \& Motivation}
\label{sec:background}

\subsection{GPU Performance Modeling}
\label{ssec:modeling_gap}

\begin{figure}[t]
    \centering
    \includegraphics[width=\linewidth]{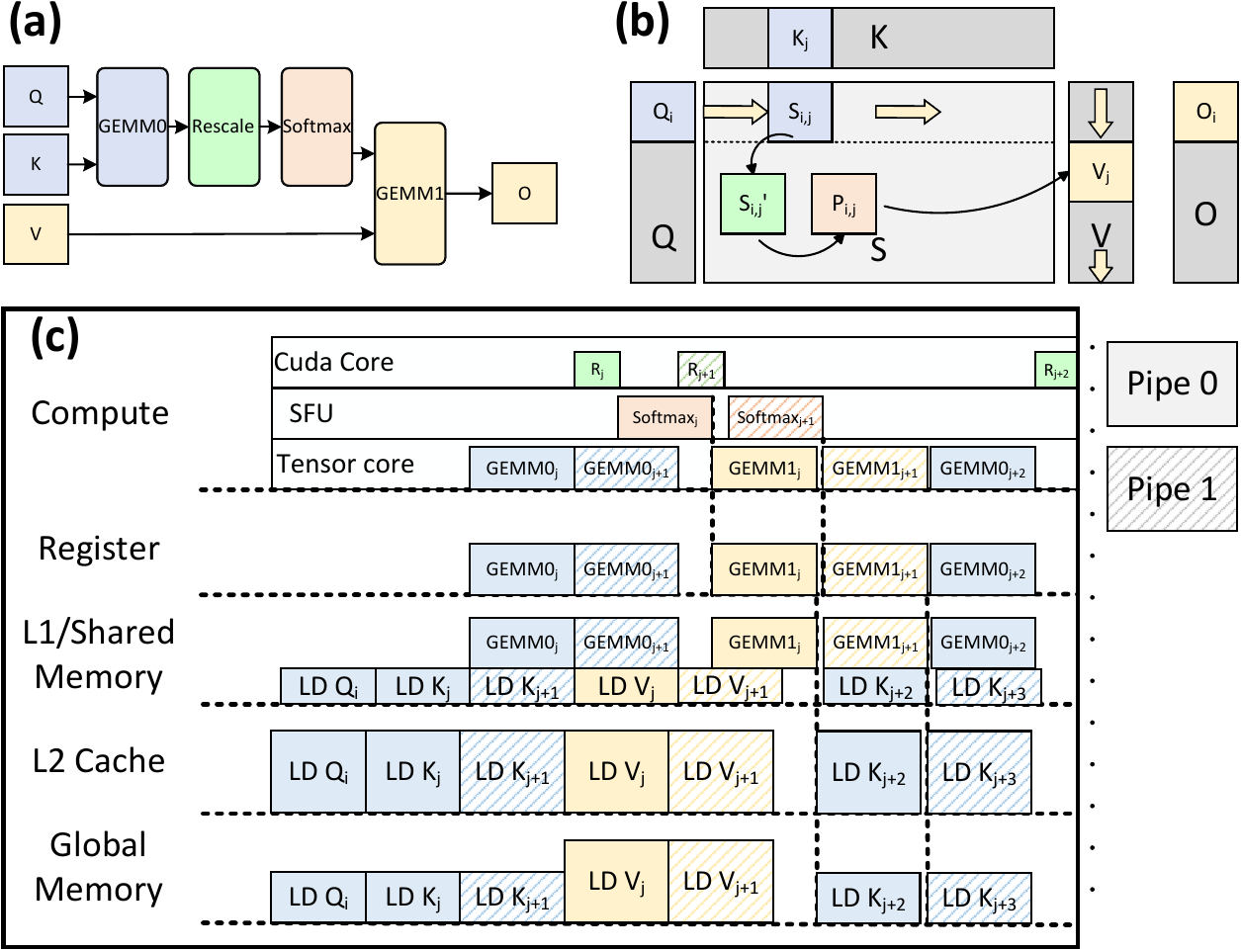}
    \caption{FlashAttention-3 on H100: (a) the 10+ heterogeneous operations spanning Tensor Cores, CUDA cores, and SFUs; (b) their data-dependency DAG; (c) how scheduling order determines compute-memory pipeline overlap.}
    \label{fig:fa3_mapping}
\end{figure}

\begin{figure}[t]
    \centering
    \includegraphics[width=0.95\linewidth]{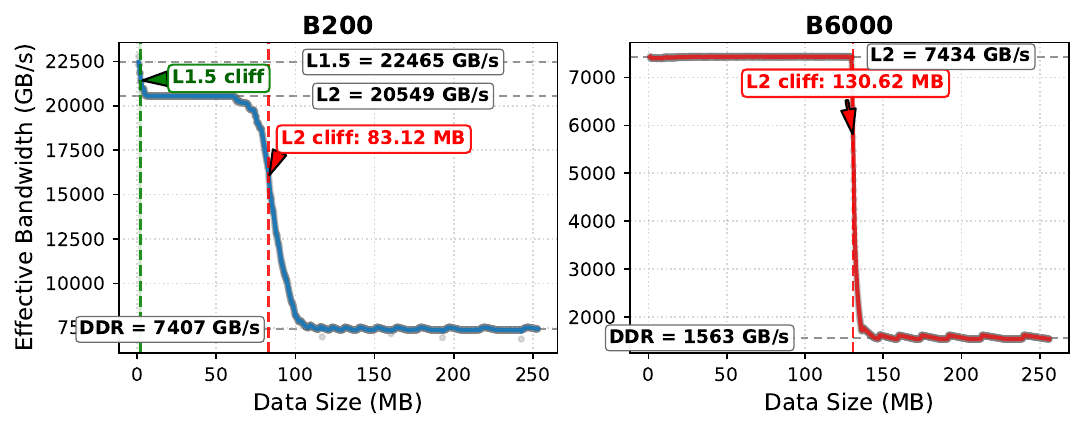}
    \caption{L2 bandwidth vs.\ working-set size on B200 and B6000, revealing the multi-level cache hierarchy. B200 (dual-die) exposes a level-1.5 (L1.5)/LRC tier at ${\sim}$22.5\,TB/s and a smeared L2 cliff at ${\sim}$83\,MB; B6000 (single-die) shows a sharp cliff at ${\sim}$130\,MB. \oursys{} uses these sweeps to calibrate effective cache capacity per GPU.}
    \label{fig:cache_bw_sweep}
\end{figure}

Existing GPU performance tools can be grouped into three categories.
\textbf{Learned and hybrid predictors}~\cite{lee2025forecasting,pipeweave2026} fit either end-to-end runtime or analytical-model residuals from per-architecture traces, arguing that pure analytical models cannot capture modern GPU interaction complexity; both nevertheless require retraining and provide limited interpretability.
\textbf{Analytical models}~\cite{williams2009roofline,parashar2019timeloop,zheng2023tileflow} are portable and explainable, but typically collapse GPU execution into aggregate compute and bandwidth terms.
\textbf{Profiling and simulation tools}~\cite{guan2025kperfir,huang2025neutrino,agrawal2024vidur,Wang2025SimAIUA} expose measured behavior after execution, but do not forecast how a tile shape, pipeline depth, or swizzle change will perform before rerunning the kernel.
As motivated in Section~\ref{sec:intro}, the shared limitation is an abstraction mismatch: these tools do not model performance at the same tile granularity used by modern GPU programs. In particular, where hybrid predictors delegate this complexity to learned components, \oursys{} shows that a first-principles tile-centric simulation can capture it while remaining fully white-box.

\subsection{Modeling Gap for Tile-Centric Programs}

The missing abstraction appears at three levels.
\textbf{Intra-tile}: each tile uses heterogeneous pipelines spanning compute, memory, and network, so a single bottleneck scalar misses the per-pipeline structure that determines overlap (Figure~\ref{fig:fa3_mapping}).
\textbf{Inter-tile}: tile dependencies determine the legal action orderings inside a fused body, and tile execution order across the grid determines cache reuse; a single flat bandwidth number is insufficient on modern GPUs (Figure~\ref{fig:cache_bw_sweep}).
\textbf{Cross-device}: partitioned tile grids exchange data through communication pipelines that must overlap with compute rather than be added as standalone times.
Table~\ref{tab:method_comparison} summarizes how existing tools miss one or more of these levels.

\begin{table}[t]
\centering
\caption{Comparison with prior performance modeling tools.}
\label{tab:method_comparison}
\footnotesize
\renewcommand{\arraystretch}{1.15}
\setlength{\tabcolsep}{2pt}
\scalebox{0.88}{
\begin{tabular}{@{} l | c c c c c c | c @{}}
\toprule
\textbf{Feature}
  & \makecell{\textbf{Roof-}\\\textbf{line}\\\cite{williams2009roofline}}
  & \makecell{\textbf{Neu-}\\\textbf{Sight}\\\cite{lee2025forecasting}}
  & \makecell{\textbf{Pipe-}\\\textbf{Weave}\\\cite{pipeweave2026}}
  & \makecell{\textbf{GenZ}\\\cite{bambhaniya2024genz}}
  & \makecell{\textbf{Vidur}\\\cite{agrawal2024vidur}}
  & \makecell{\textbf{SimAI}\\\cite{Wang2025SimAIUA}}
  & \makecell{\textbf{Tile-}\\\textbf{Sight}\\}
\\
\midrule
No kernel profiling/training\textsuperscript{1}
  & \ding{51} & \ding{55} & \ding{55} & \ding{51} & \ding{55} & \ding{55} & \ding{51} \\
\midrule
Pipeline-aware\textsuperscript{2}
  & \ding{55} & \ding{55} & \Circle & \ding{55} & \ding{55} & \ding{55} & \ding{51} \\
\midrule
Cache-aware\textsuperscript{3}
  & \ding{55} & \ding{55} & \ding{55} & \ding{55} & \ding{55} & \ding{55} & \ding{51} \\
\midrule
Explicit fused program\textsuperscript{4}
  & \ding{55} & \ding{55} & \Circle  & \ding{55} & \ding{55} & \ding{55} & \ding{51} \\
\midrule
Distributed\textsuperscript{5}
  & \ding{55} & \Circle & \Circle  & \Circle  & \ding{51} & \ding{51} & \ding{51} \\
\midrule
Compute-comm. overlap\textsuperscript{6}
  & \ding{55} & \ding{55} & \ding{55} & \ding{55} & \ding{55} & \Circle  & \ding{51} \\
\midrule
Interpretable\textsuperscript{7}
  & \ding{51} & \ding{55} & \Circle  & \ding{51} & \ding{55} & \ding{55} & \ding{51} \\
\bottomrule
\end{tabular}}

\parbox{\columnwidth}{
\scriptsize\raggedright
\ding{51}~Full support.\enspace \Circle~Partial.\enspace \ding{55}~Not supported.\\[3pt]
\textsuperscript{1}No kernel profiling/training: no kernel execution traces or ML training are required;
\oursys{} uses only one-time per-architecture microbenchmarks
(bandwidth/throughput/latency sweeps, $\sim$minutes).
\textsuperscript{2}Pipeline-aware: intra-tile DAG scheduling and compute-memory pipeline overlap.
\textsuperscript{3}Cache-aware: predicts L2/L1.5 hit rates and schedule-dependent tile locality effects.\\
\textsuperscript{4}Explicit fused program: user-described arbitrary multi-op DAG kernels (e.g., FA-3, MLA), not limited to a fixed set of supported patterns.
\textsuperscript{5}Distributed: multi-GPU collective communication modeling.\\
\textsuperscript{6}Compute-communication overlap: fused compute-communication kernels (e.g., AllGather+GEMM).
SimAI accepts user-specified overlap ratios but does not derive them analytically.
\textsuperscript{7}Interpretable: white-box model supporting bottleneck diagnosis.\\[3pt]
}
\end{table}

\section{Hierarchical Tile-Pipeline Model}
\label{sec:design}

\oursys{} treats the \emph{tile} as the first-class modeling unit and adopts a prologue--steady--epilogue pipeline envelope to  recursively apply at every level of the program.
A tile carries \emph{intra-tile} information (operation, src/dst placement, footprint, and the resources it occupies on each independently schedulable hardware pipeline) and participates in \emph{inter-tile} relationships (producer--consumer dependencies, concurrent issue, and execution order across loops, the tile grid, and waves).
A tiled workload is therefore a \emph{tile execution plan}: a graph of tiles annotated with these two kinds of information.
Distributed execution shares the same tile-based abstraction: a tile whose source or destination crosses devices simply gains a \texttt{Net} entry in its resource vector, and the same envelope still applies.

\subsection{From Workload to Tile Execution Plan}
\label{ssec:framework}

The input to \oursys{} is a high-level workload, such as a tiled GEMM, a fused attention kernel, an all-gather followed by a GEMM, or Mixture-of-Experts (MoE) routing across GPUs; the workload fixes tensors and their placements but leaves the schedule unspecified.
\oursys{} lifts it to a tile execution plan that exposes the schedule-relevant choices needed for performance modeling: tile shape, loop and reduction order, block swizzle, software-pipeline depth, resident blocks per SM, distributed partitioning, and collective implementation.
Tile-centric DSLs such as Triton and TileLang expose most of this information directly; for hand-written kernels, the same fields are supplied manually from the kernel schedule.

\begin{figure*}[t]
\centering
\includegraphics[width=\textwidth]{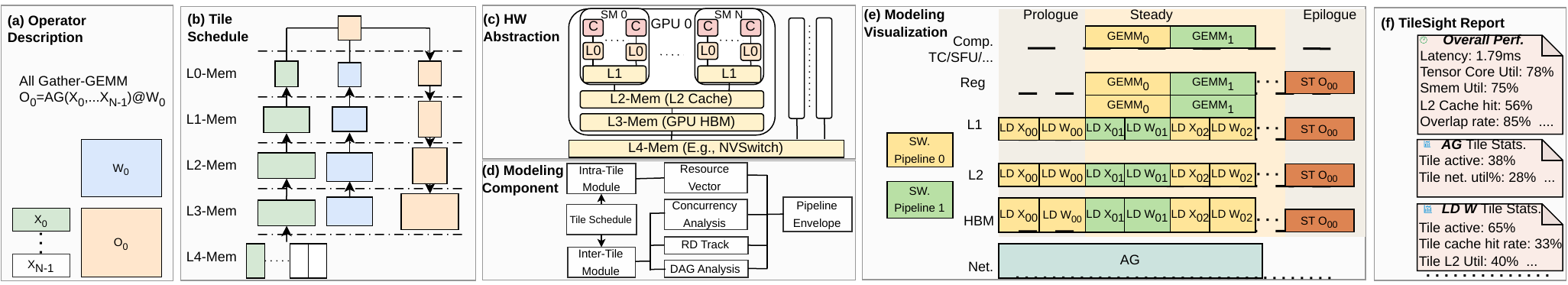}
\caption{
\textbf{\oursys{} design overview} on all-gather--GEMM (AG-GEMM). \textbf{(a)} A workload is described by an operator and tensor placement only ($X$ column-sharded across $N$ GPUs). \textbf{(b)} \oursys{} lifts it to a tile schedule whose DAG spans memory levels $L_0$--$L_4$. \textbf{(c)} A single hardware abstraction exposes registers, SMEM, L2, HBM, and the inter-GPU fabric as a 5-level hierarchy. \textbf{(d)} Intra-tile resource vectors and inter-tile DAG/concurrency analysis feed a recursive prologue--steady--epilogue envelope. \textbf{(e)} The engine renders the envelope as a timeline: software-pipelined loads overlap with compute, and the AllGather is \emph{inferred from placement} on the \texttt{Net} lane. \textbf{(f)} A per-tile performance report with latency, utilization, cache hit, and overlap rate.
}
\label{fig:hierarchy}
\end{figure*}

\begin{table}[t]
\centering
\caption{The tile execution plan groups its fields by what they describe: a single tile in isolation (intra) or relationships among tiles (inter).}
\label{tab:tile_plan_fields}
\small
\begin{tabularx}{\linewidth}{@{}>{\raggedright\arraybackslash}p{1.2cm}lX@{}}
\toprule
\textbf{Field} & \textbf{Side} & \textbf{Role in the model} \\
\midrule
Tensor accesses & Intra & Per-tile footprint and a placement descriptor recording where a tensor tile is produced and where it resides: register, architecture-specific tensor memory (TMEM), SMEM, local cache/DDR, or shard/replica on a GPU group. Reuse dimensions feed inter-tile cache modeling. \\
Operation type & Intra & The action a tile performs: load, store, tensor-core or CUDA-core matmul, reduction, exponential, rescaling, remote transfer, or fused composite. \\
Resource vector & Intra & Per-tile time on independently schedulable resources (tensor cores (TC), CUDA cores, SFU, TMEM, SMEM, L1.5, L2, DDR, Net); derived from the operation, footprint, placement, and calibrated hardware rates. \\
\midrule
Tile grid & Inter & Spatial tile shape, launch order, swizzle, loop/reduction depth, and distributed partition. Determines tile execution order, waves, and local work per device. \\
Producer--consumer DAG & Inter & Edges among tiles based on tensor production and consumption; together with placement, fixes the legal orderings within an iteration. \\
Concurrency \& depth & Inter & Software-pipeline stages, resident blocks per SM, and which tiles may issue together. Sets the effective pipeline depth. \\
\bottomrule
\end{tabularx}
\end{table}

The plan deliberately avoids thread-level details.
It keeps only the choices that tile programmers and distributed runtimes actually change, and those choices change which tiles enter the pipeline, what resources they occupy, and how they depend on or run concurrently with each other.
Cache traffic, wave effects, communication stages, and pipeline overlap are then derived rather than separately added.

\subsection{Intra-Tile: A Tile and Its Resource Vector}
\label{ssec:intra_tile}

A tile is characterized by its \emph{operation} (load, store, tensor-core or CUDA-core matmul, reduction, exponential, rescaling, remote transfer, or fused composite), its \emph{footprint} (per-tile bytes and FLOPs), and its \emph{src}/\emph{dst} \emph{placement descriptors} that record where its inputs are produced and where its output resides --- a register, architecture-specific tensor memory (TMEM), the shared-memory scratchpad, the L1.5 or L2 cache, DDR on the local device, or a shard or replica on a GPU group.
Placement is the central abstraction that lets the same intra-tile representation describe both fusion and cross-device movement: marking an intermediate output as register-, TMEM-, or SMEM-scope removes a global-memory store (fusion), while marking a load source as a remote shard turns the load into a cross-device transfer (distribution).

For each tile, \oursys{} converts these properties into a vector of times on independently schedulable hardware resources:
\begin{equation}
\mathbf{u}(o)=
\langle t_{\text{TC}}, t_{\text{CUDA}}, t_{\text{SFU}}, t_{\text{TMEM}},
t_{\text{SMEM}}, t_{\text{L1.5}}, t_{\text{L2}}, t_{\text{DDR}}, t_{\text{Net}}\rangle ,
\label{eq:usage_vector}
\end{equation}
computed from the tile's operation, footprint, src/dst placement, and one-shot microbenchmark-calibrated rates.
A pure tensor-core matmul tile populates only the TC entry; a Blackwell attention tile also charges explicit TMEM traffic for softmax and correction loads/stores; a load tile from DDR populates DDR (and L1.5/L2 if the access hits cache); a remote-load tile populates Net.
This vector is more expressive than a roofline scalar because tiles on different pipelines may overlap, while tiles contending for the same pipeline serialize, and remote movement composes with local compute through the same machinery.
Two entries of $\mathbf{u}(o)$ are not fixed by the tile in isolation: the L1.5/L2/DDR split for a memory tile depends on whether its access hits cache, derived from tile reuse distance in \S\ref{ssec:cache_model}; the \texttt{Net} entry for a remote tile depends on the routed cost of the underlying communication stage, derived in \S\ref{ssec:distributed}.
Algorithm~\ref{alg:pipeline_overlap} sketches how all components plug into the master loop; subsequent subsections detail each block.

\begin{algorithm}[t]
\caption{\footnotesize Hierarchical Tile-Pipeline Evaluation }
\label{alg:pipeline_overlap}
\scriptsize
\DontPrintSemicolon
\SetKwInOut{Input}{Input}
\SetKwInOut{Output}{Output}
\Input{tile execution plan $P$, hardware specification $H$, optional distributed mapping $\Pi$}
\Output{predicted latency $T$ and per-resource utilization}

$G \leftarrow$ tile grid, launch order, and swizzle from $P$\;
$A \leftarrow$ tensor accesses, reuse dimensions, and placement descriptors from $P$\;
$D \leftarrow$ tile-action DAG from $P$\;
$S \leftarrow$ software-pipeline parameters from $P$\;

\If{$\Pi$ is not empty}{
    $G,A,D \leftarrow \textsc{PartitionTilePlan}(G,A,D,\Pi)$\tcp*{single device is the local-only case}
    $\mathcal{O}_{net} \leftarrow \textsc{InferRemoteTensorAccesses}(G,A,D,\Pi)$\;
    $\mathcal{N} \leftarrow$ network topology and calibrated $\alpha,\beta$ parameters from $H$\;

    \ForEach{remote tensor access sequence $c \in \mathcal{O}_{net}$}{
        $\mathcal{K}_c \leftarrow \textsc{DecomposeIntoStages}(c)$\tcp*{e.g., ring steps or tree levels}
        \ForEach{stage $k \in \mathcal{K}_c$}{
            $\mathcal{E}_k \leftarrow \textsc{LogicalExchanges}(k)$\tcp*{tuples $(src,dst,bytes)$}
            $\mathcal{R}_k \leftarrow \textsc{Route}(\mathcal{E}_k,\mathcal{N})$\;
            $T_k, U_k \leftarrow \textsc{AlphaBetaStageTime}(\mathcal{R}_k,\mathcal{N})$\;
        }
        Annotate the corresponding transfer tile in $D$ with $\sum_k T_k$ on \texttt{Net}\;
    }
}

$C \leftarrow \textsc{CacheTraffic}(G,A,H)$\;
Annotate memory tiles in $D$ with L1.5/L2/DDR entries from $C$\;

$p \leftarrow \textsc{ResidentTilesPerSM}(P,H)$\;
$d \leftarrow S.\text{stages}\times p - 1$\;
$\mathcal{E}_{tile} \leftarrow \textsc{PipelineEnvelope}(D,d,H,\text{active SMs})$\;
$T,\ U \leftarrow \textsc{WaveAggregate}(G,\mathcal{E}_{tile},H)$\;
\Return $T,\ U$\;
\end{algorithm}

\subsection{Inter-Tile: Dependency, Concurrency and Order}
\label{ssec:inter_tile}

Tiles connect through three kinds of inter-tile information.
\textit{1)} \emph{Producer--consumer dependencies} fix the legal orderings within an iteration: in FlashAttention, $Q$/$K$ loads precede gemm1 ($Q\!@\!K$), gemm1 precedes softmax, and softmax precedes gemm2 ($P\!@\!V$). Together with placement and dependencies, it determines which intermediates stay in registers/TMEM/SMEM and which spill to global memory.
\textit{2)} \emph{Concurrent issue} lets non-dependent tiles run together when their resource vectors do not contend, e.g., concurrently loading the next $K$-block of attention while computing on the current one, or issuing the A and B loads of a GEMM along the same $K$ slice. The same set of tiles can be ordered in multiple legal ways that yield different overlap on shared pipelines.
\textit{3)} \emph{Tile execution order} across loop iterations and the tile grid determines which loads find their data already resident in cache: row-panel traversal preserves B-tile reuse for adjacent $M$-rows, block swizzle reorders the sequence, and persistent-block schedules pin tiles to SMs.
These three pieces are exactly the input the pipeline envelope needs.

\subsection{Pipeline Envelope: Prologue--Steady--Epilogue}
\label{ssec:pse}

Given a set of tiles with resource vectors and inter-tile relationships, \oursys{} evaluates execution as a pipeline.
For a repeated unit with $N$ logical iterations and effective depth $d$:
\begin{equation}
T =
T_{\text{pro}} +
\max(N-d,0)\,T_{\text{steady}} +
T_{\text{epi}},
\label{eq:pipeline_envelope}
\end{equation}
where $T_{\text{pro}}$ is the fill cost, $T_{\text{steady}}$ is the overlapped cost per repeated unit, and $T_{\text{epi}}$ is the drain cost.
The same envelope applies recursively at every level of the tile execution plan: the steady-state body of an outer envelope (over tile-block waves) can itself be a pipeline (over a $K$-loop), whose steady body can in turn be a pipeline over the inner action sequence.
The effective depth combines explicit software-pipeline stages with resident tile interleaving:
\begin{equation}
d = \text{stages} \times \text{resident\_tiles\_per\_SM} - 1 .
\label{eq:effective_depth}
\end{equation}
A two-block-per-SM schedule is therefore not a special case: it deepens the pipeline because an SM can issue work from one resident tile-block while another waits on memory.

\para{Steady-state overlap.}
The steady-state cost of a tile sequence depends on which legal ordering is chosen, since tiles using the same hardware dimension in Eq.~\ref{eq:usage_vector} accumulate on that dimension while independent dimensions overlap:
\begin{equation}
T_{\text{steady}}(\sigma)
=
\max_{r}
\sum_{o \in \sigma} u_r(o),
\label{eq:steady_overlap}
\end{equation}
subject to all data-dependency edges in the DAG.
The selected steady state is the best legal ordering:
\begin{equation}
T_{\text{steady}} =
\min_{\sigma \in \textsc{Topo}(D)} T_{\text{steady}}(\sigma).
\label{eq:topo_select}
\end{equation}
This is a small search in practice because real fused-kernel DAGs are heavily constrained.
For MLA decode, 11 tile actions reduce from $11!$ unconstrained permutations to 132 legal topological orders.
The search is not an autotuning run: it is an analytical scheduling step over the tile plan, so it remains cheap enough to run inside a cost model.

\para{Boundary costs.}
The prologue and epilogue are computed from the same resource vectors but with reduced overlap.
For a load--compute pipeline, the prologue consists primarily of memory tiles that fill the pipeline, while the epilogue consists of remaining compute and final stores.
Fused tile bodies add reductions or normalizations to one or both boundaries.
This separation matters because two schedules with the same steady-state bottleneck can have different end-to-end time when the loop count is short or when only a few waves are launched.

\para{Resident tiles and waves.}
Occupancy changes overlap structure, not only utilization.
If $p$ tile-blocks reside on one SM, the model treats them as interleaved instances of the same tile pipeline; the resident count is bounded by shared memory, registers, warp limits, and architecture-specific maximum blocks per SM.
The same wave decomposition handles tail effects: a tail wave may use only a subset of SMs, and those active SMs receive a larger share of shared L2/DDR bandwidth, so the envelope is recomputed for the tail using its active-SM count.
Algorithm~\ref{alg:tile_overlap} expands this evaluation, recursively traversing the tile loop structure and enumerating dependency-valid orderings.

\begin{algorithm}[t]
\caption{\footnotesize Recursive Pipeline-Envelope Evaluation}
\label{alg:tile_overlap}
\scriptsize
\SetAlgoLined
\DontPrintSemicolon
\SetKwFunction{OverlapAnalysis}{OverlapAnalysis}
\SetKwFunction{AnalyzeLoop}{AnalyzeLoop}
\SetKwFunction{ModelOverlap}{ModelOverlap}
\SetKwFunction{ComputeTilesPerSM}{ResidentTilesPerSM}
\SetKwFunction{WaveDecompose}{WaveDecompose}
\SetKwFunction{GetSubNodes}{GetSubNodes}
\SetKwFunction{GetPipelineStage}{GetPipelineStage}
\SetKwFunction{MergeMetrics}{MergeMetrics}
\SetKwProg{Fn}{Function}{:}{end}

\Fn{\OverlapAnalysis{$P,H$}}{
  $p \gets \ComputeTilesPerSM(P,H)$\;
  $(n_{\mathrm{full}}, n_{\mathrm{tail}}) \gets \WaveDecompose(P.\mathrm{grid}, H.\mathrm{SMs}, p)$\;
  $(T^{\mathrm{full}}, U^{\mathrm{full}}) \gets \AnalyzeLoop(P.\mathrm{root}, p, H.\mathrm{SMs})$\;
  \If{$n_{\mathrm{tail}} > 0$}{
    $(T^{\mathrm{tail}}, U^{\mathrm{tail}}) \gets \AnalyzeLoop(P.\mathrm{root}, p, n_{\mathrm{tail}})$\;
  }
  \Else{
    $T^{\mathrm{tail}}\gets 0,\ U^{\mathrm{tail}}\gets \emptyset$\;
  }
  \Return $n_{\mathrm{full}}T^{\mathrm{full}} + T^{\mathrm{tail}},\ \MergeMetrics(U^{\mathrm{full}},U^{\mathrm{tail}})$\;
}

\BlankLine
\Fn{\AnalyzeLoop{$node, stage, active\_SMs$}}{
  $groups \gets \GetSubNodes(node)$\;
  \If{$node$ is an inner loop}{
    $s \gets \GetPipelineStage(node)$\;
    $d \gets s \times stage - 1$\tcp*{software stages $\times$ resident tiles}
    \Return \ModelOverlap($groups,d,active\_SMs$)\;
  }
  $metrics \gets [\,]$\;
  \ForEach{$g \in groups$}{
    \If{$g$ is a loop}{
      $metrics.\mathrm{append}(\AnalyzeLoop(g,stage,active\_SMs))$\;
    }
    \Else{
      $metrics.\mathrm{append}(\ModelOverlap([g],stage-1,active\_SMs))$\;
    }
  }
  \Return \MergeMetrics($metrics$)\;
}

\BlankLine
\Fn{\ModelOverlap{$groups,d,active\_SMs$}}{
  $N \gets$ repeated-iteration count represented by $groups$\;
  $best \gets \infty$\;
  \ForEach{$\sigma \in \textsc{Topo}(groups)$}{
    $\mathbf{u}_{\sigma} \gets$ resource-vector accumulation under order $\sigma$ and $active\_SMs$\;
    $T_{\mathrm{pro}},T_{\mathrm{steady}},T_{\mathrm{epi}} \gets$ boundary and steady costs from $\mathbf{u}_{\sigma}$\;
    $T \gets T_{\mathrm{pro}}+\max(N-d,0)T_{\mathrm{steady}}+T_{\mathrm{epi}}$\;
    \If{$T < best$}{ $best \gets T$ \; }
  }
  \Return $best$ and the corresponding utilization\;
}
\end{algorithm}

\subsection{Cache Traffic via Tile Reuse Distance}
\label{ssec:cache_model}

For a memory tile, the L1.5/L2/DDR split is not a property of the tile in isolation: the same load-tile coordinate can hit cache or fall through to DDR depending on swizzle, wave occupancy, and which neighboring tiles share tensor data.
Preserving B-tile reuse across GEMM $M$-axis tiles can cut DDR traffic by ${\sim}4\times$, and block swizzling shifts L2 hit rate from 35\% to 72\% in our motivating case; modern GPUs further add intermediate L1.5/LRC tiers (H200, B200), making a single flat bandwidth term insufficient.
Reuse-distance analysis is well established for cache modeling~\cite{lam1991cache,conte1998combining,nugteren2014detailed,arafa2019gpus,arafa2020fast,niu2012parda}, but conventional formulations operate on cache-line traces and are too low-level to place inside analytical schedule search.
\oursys{} instead lifts reuse distance to the tile execution plan, with the symbolic tile order as the analyzed sequence and tile-sized tensor blocks as the reuse universe --- to our knowledge, the first analytical GPU performance model to make schedule-sensitive, multi-level cache modeling practical through a tile-granular reuse-distance abstraction.

\subsubsection{Tensor Access and Tile Reuse Distance}
\label{sssec:reuse_distance}

\oursys{} introduces a \emph{tensor access} for each tensor associated with the tile grid: per-tile footprint, placement descriptor, repeated-access count, and the grid dimensions along which the same data block is reused.
The reuse dimensions \texttt{reuse\_dims} make one rule cover diverse operators: a tensor's reuse key is the tile coordinate projected onto the non-reuse dimensions.
For a GEMM grid $(M_t,N_t)$, A tiles are reused across $N_t$ and B tiles across $M_t$.
For MLA decode, key--value (KV)-cache tiles are reused across attention heads of the same batch element.
For convolution, weights and activations have different reuse dimensions over batch, output-channel, and spatial axes.
This avoids operator-specific cache formulas while preserving the schedule information that determines reuse.

The \emph{tile reuse distance} $D_T$ is the number of distinct tile-sized data blocks accessed between two consecutive accesses to the same tensor block.
Traditional reuse distance asks how many cache lines or memory transactions intervene between two accesses, tile reuse distance asks the same question at the unit GPU kernel schedules expose.
Modeling an 8~KB tile instead of 128-byte cache lines reduces tracked entries by $64\times$, matches the granularity tile-centric schedules expose, makes block swizzles and traversal orders directly visible to the cache model, and avoids trace-level cache simulation.

\begin{figure}[t]
\centering
\includegraphics[width=\columnwidth]{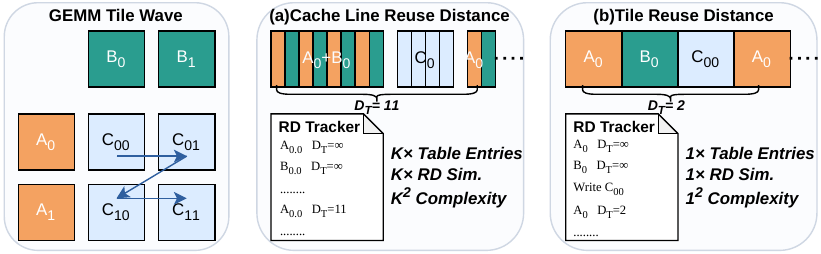}
\caption{%
\textbf{Tile vs.\ cache-line reuse distance.}
\textit{Left}: traditional cache-line reuse distance tracks tens of thousands of line entries and evaluates exact SDCM at line granularity.
\textit{Right}: \oursys{} lifts reuse distance to tile-sized blocks, applies a Gaussian SDCM approximation, and samples along reduction axes, preserving schedule sensitivity while making cache modeling lightweight.}
\label{fig:tile_reuse_distance}
\end{figure}

For a tensor with \texttt{reuse\_dims}, \oursys{} computes a reuse key from the tile's non-reuse coordinates:
\begin{equation}
\text{key}(\mathbf{x}, R)=\textsc{Linearize}\bigl(x_d\;|\;d\notin R\bigr),
\label{eq:reuse_key}
\end{equation}
where $\mathbf{x}$ is the tile coordinate and $R$ is the set of reuse dimensions.
For GEMM's A matrix with $R=\{N_t\}$, all tiles in the same M-row share the same A key. 
For B, all tiles in the same N-column share the same B key.
The concrete tile execution order, including swizzles and row-panel traversal, determines the sequence in which these keys appear and therefore their reuse distances.

\subsubsection{Hit Probability and Fast Evaluation}
\label{sssec:hit_probability}

Given reuse distance $D_T$, associativity $A$, and cache capacity $B_T$ measured in tile units, the stochastic distance cache model estimates the hit probability of a least-recently-used (LRU)-like cache.
The exact SDCM hit probability can be expressed with a binomial form:
\begin{equation}
P(h \mid D_T) =
\sum_{a=0}^{A-1}
\binom{D_T}{a}
\left(\frac{A}{B_T}\right)^a
\left(\frac{B_T-A}{B_T}\right)^{D_T-a},
\label{eq:sdcm}
\end{equation}
where $A$ is cache associativity and $B_T$ is cache capacity measured in tiles.
While accurate, this binomial form is expensive to compute for every tensor key in a large tile grid.

\oursys{} therefore adopts a Gaussian approximation for efficient evaluation:
\begin{equation}
P(h \mid D_T)_{\text{approx}}
=
1 - Q\!\left(
\frac{|A-1-\mu|}{\sqrt{\sigma^2}}
\right),
\label{eq:sdcm_gaussian}
\end{equation}
where
\begin{equation}
\mu = D_T \cdot \frac{A}{B_T},
\qquad
\sigma^2 =
D_T \cdot \frac{A}{B_T}
\cdot
\left(1-\frac{A}{B_T}\right).
\label{eq:sdcm_moments}
\end{equation}
$Q(x)$ denotes the complementary cumulative distribution function (CDF) of the standard normal distribution.
To further reduce overhead, we apply the Zelen--Severo approximation~\cite{abramowitz1965handbook} for the CDF $\Phi(x)$:
\begin{equation}
\Phi(x)
\approx
1 -
\left(a_1t-a_2t^2+a_3t^3\right)
\frac{e^{-x^2/2}}{\sqrt{2\pi}},
\label{eq:zelen_severo}
\end{equation}
where $t=(1+0.33267x)^{-1}$ and $a_1,a_2,a_3$ are constants.

\para{Sampling along reduction axes.}
Tile execution plans expose reduction axes (e.g., the $K$ axis in GEMM). \oursys{} samples reuse events at this granularity rather than replaying every inner-loop access (a GEMM with $K{=}8192$, $\text{tile}_K{=}32$ reduces checks by $256\times$ with negligible accuracy loss).
Together with tile-level reuse distance and the Gaussian approximation, this reduces cache-model evaluation by roughly five orders of magnitude, enabling cache modeling inside the analytical loop rather than as offline trace analysis.

\subsubsection{Two-Level Cascade, Swizzle, and Waves}
\label{sssec:sdcm_cascade}

On GPUs with an intermediate L1.5/LRC tier, \oursys{} applies SDCM as a cascade --- L1.5 within each physical SM group, L2 globally, and DDR carrying the residual miss traffic; without this design, L1.5 hit probability is zero and the model collapses to a single L2 evaluation.
A block swizzle, row-panel, Z-order, or persistent-block schedule is just a concrete sequence of tile coordinates fed to the reuse-distance simulation.
Within a wave, \oursys{} perturbs $D_T$ for hardware nondeterminism, sequential tensor loads, and cross-tensor cache aging, all derived from the tile execution plan and hardware grouping with no kernel-specific profiling.
Tail waves use a subset of SMs and therefore receive a larger share of shared bandwidth, so the envelope is recomputed for the tail.
The resulting L1.5/L2/DDR byte counts populate the corresponding entries of Eq.~\ref{eq:usage_vector}, so cache behavior changes the pipeline envelope itself, not only the final latency.

\subsection{Cross-Device Tiles}
\label{ssec:distributed}

Cross-device execution is a placement extension of the same intra-tile abstraction: a tile's source or destination can point to a shard or replica on another GPU, and its resource vector picks up a non-zero \texttt{Net} entry.
A tensor-, expert-, sequence- or data-parallel mapping partitions both the tile grid and its tensor tiles, producing placement descriptors over GPU groups.
After partitioning, a local tile wave may need a tensor tile produced by another device, a replicated activation, or a partial result that must be reduced before later tiles can consume it.
\oursys{} treats these as \emph{remote tensor accesses}: the required collectives or point-to-point transfers are inferred directly from producer--consumer placement, and each becomes a tile with source/destination devices, byte volume, and \texttt{Net} resource usage.

\para{Logical exchanges and topology.}
For each inferred remote tensor access, \oursys{} decomposes the required tensor-tile movement into ordered stages.
A stage is represented by logical source--destination exchanges $(s,d,b)$, where $s$ is the device that owns or produces the tensor tile, $d$ denotes the device whose tile wave consumes it, and $b$ means the tile or shard byte volume derived from the tensor access.
Collective algorithms simply provide different stage decompositions: a ring all-reduce uses reduce-scatter and all-gather steps, tree algorithms use reduction and broadcast levels, and irregular routing remains point-to-point.
This representation is tile-level rather than packet-level.
It preserves the tensor-placement information needed to reason about communication volume, while leaving the hardware topology to determine which physical network-on-chip (NoC) or interconnect links carry each exchange.

\para{Per-stage routed cost.}
After routing the exchanges in a stage, \oursys{} estimates the stage time with an $\alpha$--$\beta$ communication model~\cite{thakur2005optimization} that matches the decomposition into hop latency and bottleneck-link serialization:
\begin{equation}
T_k
=
\underbrace{
\max_{(s,d,b)\in\mathcal{E}_k}
\sum_{l\in\mathcal{P}_{sd}} \alpha_l
}_{\text{routed hop latency}}
+
\underbrace{
\max_{l\in\mathcal{L}} \beta_l B_{l,k}
}_{\text{bottleneck-link serialization}},
\label{eq:stage_alphabeta}
\end{equation}
where $\mathcal{E}_k$ is the set of logical exchanges in stage $k$, $\mathcal{P}_{sd}$ the physical route for $(s,d,b)$, $B_{l,k}$ the bytes routed through link $l$, and $\alpha_l,\beta_l$ are the calibrated startup latency and inverse bandwidth of link $l$.
The cost of an inferred communication sequence is the ordered sum over its stages, $T_c=\sum_{k\in\mathcal{K}_c}T_k$.
For algorithms with repeated identical stages such as ring collectives, \oursys{} evaluates one stage and multiplies by the stage count.
The result enters the \texttt{Net} dimension of Eq.~\ref{eq:usage_vector}, so cross-device movement is represented as an intra-tile resource requirement and overlaps with local compute through the same steady-state machinery as everything else.

\subsection{Putting It Together}
\label{ssec:closure}

With the pieces in place, Algorithm~\ref{alg:pipeline_overlap} regains its full meaning: cache analysis (\S\ref{ssec:cache_model}) populates the L1.5/L2/DDR entries of $\mathbf{u}(o)$ from inter-tile execution order; remote tensor accesses (\S\ref{ssec:distributed}) populate the \texttt{Net} entry from routed $\alpha$--$\beta$ stage cost; the envelope (\S\ref{ssec:pse}) then consumes the completed resource vectors and the dependency/concurrency edges (\S\ref{ssec:inter_tile}), applied recursively across nested loops, waves, and network stages.
None of these are post-hoc corrections --- each piece either populates or consumes the same per-tile resource vector that flows through the envelope.

\subsection{Portable Hardware Abstraction}
\label{ssec:hw_abstraction}

\oursys{} requires only the parameters that affect the tile execution plan and its placement descriptors.
The abstraction mirrors the tensor-placement hierarchy: local placements map to register/TMEM/SMEM/cache/DDR resources, remote placements map to a calibrated network hierarchy across GPUs and nodes (Table~\ref{tab:gpu_specs}).
Values come from vendor specifications and lightweight microbenchmarks for practical bandwidth, utilization caps, 
and network parameters.

\begin{table}[t]
\centering
\caption{Hardware specifications: theoretical peak (spec) / microbenchmark-calibrated (meas.) for GPU architectures evaluated in this paper.}
\label{tab:gpu_specs}
\resizebox{\linewidth}{!}{%
\begin{tabular}{@{}lrllllr@{}}
\toprule
\textbf{GPU} & \textbf{SMs} & \textbf{VEC FP32 T} & \textbf{TC FP16 T} & \textbf{SFU T} & \textbf{L2 TB/s} & \textbf{DDR TB/s} \\
             &              & \footnotesize{spec~/~meas.} & \footnotesize{spec~/~meas.} & \footnotesize{spec~/~meas.} & \footnotesize{meas.} & \footnotesize{spec~/~meas.} \\
\midrule
A100  & 108 & 19.5~/~19.0  & 312~/~299   & 2.4~/~2.4   & 3.2  & 1.9~/~1.7  \\
H200\textsuperscript{*}  & 132 & 61.8~/~49.5  & 989~/~928   & 3.9~/~4.1   & 9.2 & 4.8~/~4.2  \\
B6000 & 188 & 117~/~88.6   & 468~/~433   & 7.3~/~6.7  & 7.6  & 1.8~/~1.4  \\
B200  & 148 & 74.5~/~57.7  & 2382~/~2185 & 4.7~/~4.5   & 20.5 & 8.0~/~7.0  \\
MI210 & 104 & 45.3~/~34.4  & 181~/~167   & 2.8~/~1.1   & 4.8  & 1.6~/~1.4  \\
\bottomrule
\end{tabular}}

\parbox{\columnwidth}{
\scriptsize
\textit{Note}: \oursys{}'s hardware abstraction also includes cache hierarchy, architecture-specific TMEM bandwidth, SMEM/occupancy limits, and network hierarchy across GPU groups. Not listed here for simplicity.
\textsuperscript{*}H200 has a maximum clock of 1980\,MHz and a default clock of 1830\,MHz.
}
\end{table}

\oursys{} does not model warp-level instruction issue, compiler register allocation, hardware scheduling at instruction granularity, or packet-level network effects.
Instead, it models the schedule-visible effects that tile-level programmers and distributed runtimes control: tile shape, tensor placement, reuse pattern, swizzle order, pipeline depth, resident blocks per SM, distributed partitioning, collective algorithm, and topology-aware routing.
This is what makes the model both portable across GPU generations and fast enough to use inside schedule search.

\section{Implementation}
\label{sec:implementation}

\oursys{} is implemented in Python ($\sim$6K lines) and supports NVIDIA and AMD GPUs.
Users describe kernels as tile-based programs, either extracted from Triton or TileLang code, or written by hand for non-DSL kernels, and \oursys{} produces a full performance breakdown without running the kernel.

\para{Describing arbitrary fused programs.}
To represent arbitrary kernels, \oursys{} describes the operations executed within each tile as a tile-action DAG (\S\ref{ssec:framework}).
Each tile action is annotated with the \texttt{HardwareUsage} resource vector~(\secref{ssec:framework}) and two additional attributes: (1) explicit data dependencies among actions, and (2) the \emph{scratchpad memory level at which intermediate results reside}: the register file, shared memory, or architecture-specific tensor memory (TMEM) on Blackwell.
The scratchpad annotation determines the bandwidth tier charged for each data movement between actions and how much on-chip capacity is consumed, which in turn constrains occupancy.
Data dependencies are declared between tile-action nodes.
\oursys{} automatically enumerates all valid topological orderings consistent with these dependencies and selects the schedule minimizing tile latency.

\para{Software pipeline and occupancy.}
For pipelined kernels, the user provides the pipeline depth, corresponding to \texttt{num\_stages} in Triton or explicit stage counts in TileLang.
Given kernel resource usage, such as shared memory per tile and register count, \oursys{} computes the number of resident tiles per SM as the resource-limited minimum.
This determines the effective pipeline depth and per-SM bandwidth allocation.
\oursys{} models head and tail waves separately: the tail wave has fewer active SMs, so each SM has a larger per-SM share of L2 and DDR bandwidth, which is reflected in the per-tile latency computation.

\para{Single GPU to cluster.}
At the single-GPU level, the entire tile grid is scheduled on one device.
At the node level, a \texttt{DistributedTileMap} partitions the grid across GPUs and a \texttt{NetworkHierarchy} captures the intra-node interconnect, including NVLink or PCIe. \oursys{} selects collective algorithms, such as ring, recursive-doubling, Rabenseifner, based on message size and device count.
For multi-node clusters, the same \texttt{NetworkHierarchy} is extended with inter-node links, such as InfiniBand or NVLink Bridge.
Users can specify custom topologies by providing per-hop bandwidth and latency for any link.
Given a \texttt{DistributedTileMap}, \oursys{} infers the required remote tensor accesses from producer--consumer placement of the partitioned tile grid, decomposes each into ordered stages of $(s,d,b)$ logical exchanges, and applies the $\alpha$--$\beta$ stage cost over the \texttt{NetworkHierarchy} to produce a per-tile \texttt{Net} resource time that flows through the same pipeline envelope as local compute and memory.

\begin{figure*}[t]
    \centering
    \includegraphics[width=\textwidth]{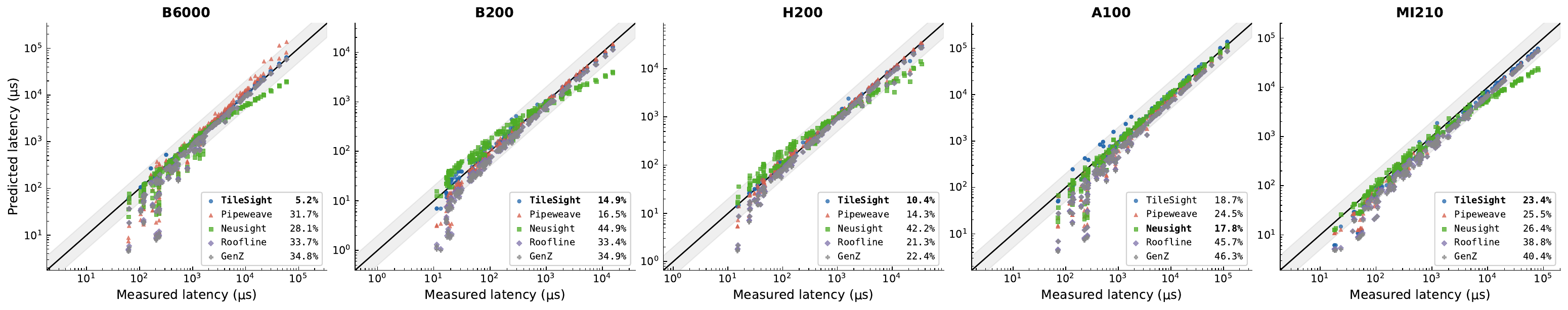}
    \caption{GEMM latency prediction vs.\ measured latency across A100, B200, B6000, H200, and MI210. Each point is one BF16/FP16 tensor-core GEMM shape; the diagonal indicates exact prediction.}
    \label{fig:gemm_cross_scatter}
\end{figure*}

\begin{table}[t]
\centering
\scriptsize
\caption{FlashAttention-3 modeling compared with NCU on H100 (Qwen configuration: batch 1, 64 heads, head-dim 128). NCU is ground truth.}
\label{tab:fa3_final_comparison}
\setlength{\tabcolsep}{2.5pt}
\begin{tabular}{@{}lrrrrrr@{}}
\toprule
\textbf{} & \textbf{Time (ms)} & \textbf{L2 hit (\%)} & \textbf{L2 util. (\%)} & \textbf{SMEM (\%)} & \textbf{TC (\%)} & \textbf{SFU (\%)} \\
\midrule
NCU & 5.58 & 96.50 & 38.66 & 51.14 & 74.78 & 38.58 \\
\oursys{} & 5.73 & 95.26 & 35.72 & 43.13 & 70.30 & 35.42 \\
\bottomrule
\end{tabular}
\end{table}

\para{Composition and calibration.}
The modeling chain runs bottom-up: our cache model computes L1.5/L2/DDR traffic fractions based on tile schedule and reuse distances.
These traffic data feed into the per-tile pipeline overlap model.
The wave model aggregates per-tile results into per-device time, while the distributed model adds communication and computes overlap.
Memory/TMEM bandwidth, per-unit compute throughput, and other hardware parameters are calibrated once per architecture with small microbenchmarks.
This consists of bandwidth sweeps over working-set sizes, as in Figure~\ref{fig:cache_bw_sweep}, and short matrix-multiply probes that only take seconds.

\begin{figure}[t]
    \centering
    \includegraphics[width=\linewidth]{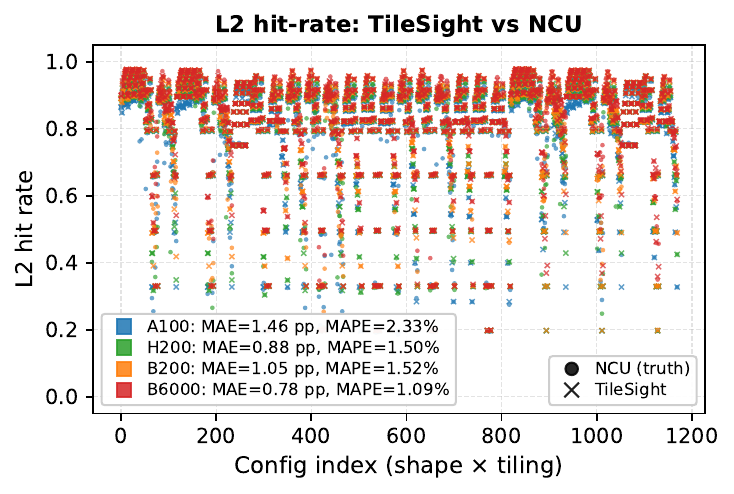}
    \caption{\oursys{} L2 hit-rate prediction vs.\ NCU ground truth across 4{,}680 GEMM persistent-kernel cases. 
    }
    \label{fig:cache_hit_eval}
\end{figure}

\section{Evaluation}
\label{sec:eval}

\begin{figure*}[t]
    \centering
    \includegraphics[width=0.98\textwidth]{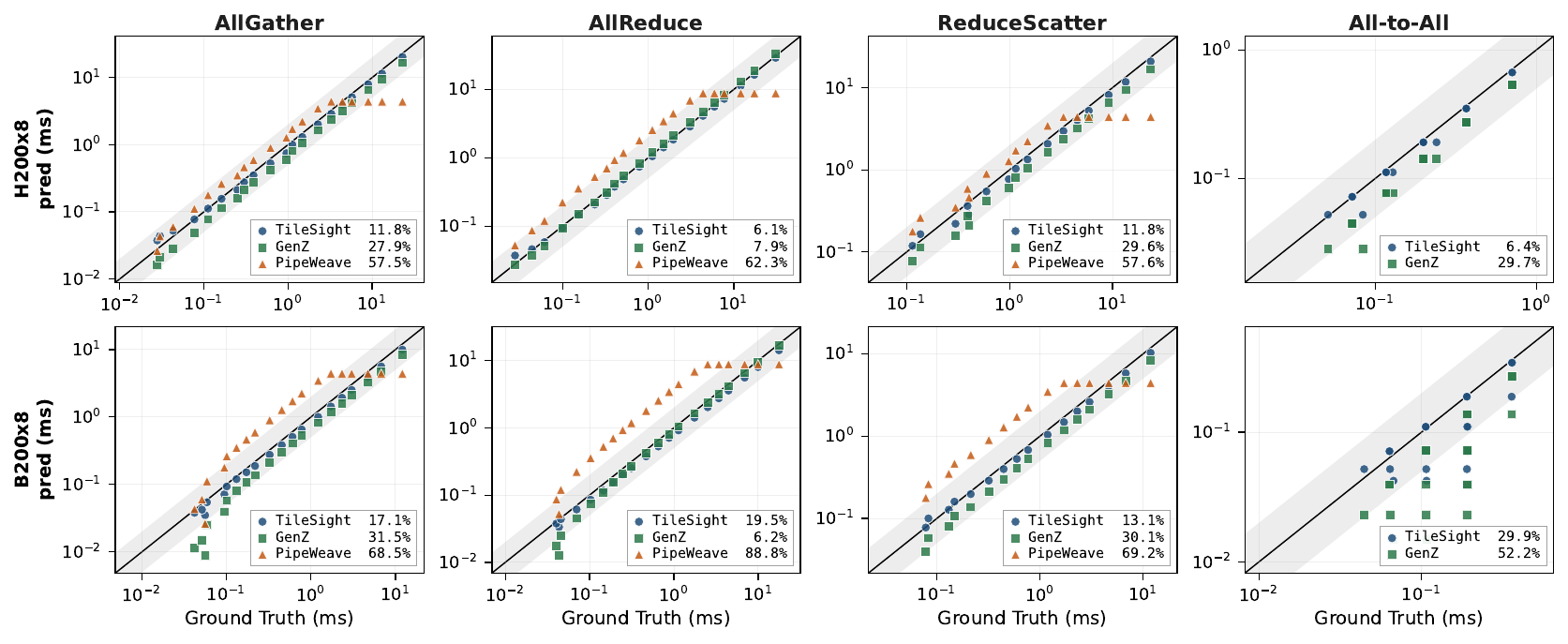}
    \caption{Pure-collective prediction on H200$\times$8 and B200$\times$8 across AllGather, AllReduce, ReduceScatter, and All-to-All.}
    \label{fig:dist_kernel_collective}
\end{figure*}

To demonstrate that a unified tile-level abstraction can model single-operator latency, cache behavior, distributed kernels, and end-to-end serving without per-target retraining or profiling, 
we evaluate \oursys{} from single-GPU kernels to multi-GPU LLM serving across A100, H200, B200, B6000, H200-NVL, and B200$\times$32 systems.

We first describe the experimental setup in \S\ref{ssec:setup}, including the hardware and framework configuration, workloads, and baselines.
We then validate the core tile-level model on single-GPU operators in \S\ref{sec:model-accuracy}, followed by a deeper analysis of L2 cache prediction for persistent kernels in \S\ref{ssec:l2_deep_dive}.
\S\ref{ssec:dist_eval} and \S\ref{ssec:e2e_cases} extend the evaluation to distributed settings, covering both collective/fused compute-communication kernels and end-to-end vLLM serving.
Finally, \S\ref{ssec:dsl_integration} shows how \oursys{} can be used for performance diagnosis and cost-model-guided schedule pruning.

\subsection{Experimental Setup}
\label{ssec:setup}

\para{Hardware and framework.}
To ensure broad hardware coverage, we evaluate A100$\times$1, H200-SXM$\times$8, B200$\times$8, an InfiniBand-connected B200$\times$32 cluster, B6000$\times$2, and H200-NVL$\times$8 systems, spanning SXM, PCIe, NVLink4/5, and multi-node settings.
We use CUDA 12.9 on Hopper and Ampere machines and CUDA 13.1 on Blackwell machines.
We additionally evaluate AMD MI210 (CDNA2) on ROCm~6.2 for cross-vendor coverage.
GEMM measurements use \texttt{cutlass\_profiler} on NVIDIA GPUs and Composable Kernel (CK) on MI210; distributed kernels use Parallel Kittens~\cite{sul2025parallelkittens}; and end-to-end serving uses vLLM 0.19.0.

\para{Workloads.}
Kernel-level experiments cover BF16/FP16 GEMMs, persistent-kernel cache sweeps, collectives, and fused compute-communication kernels.
End-to-end vLLM experiments include dense and MoE models from the Qwen, Llama, and DeepSeek families, ranging from single-GPU serving to tensor-, expert-, and data-parallel serving on up to 32 GPUs.
In total, we evaluate 703 GEMM shapes, 4{,}680 persistent-kernel cases for cache modeling, and 166 vLLM decode configurations.

\para{Baselines.}
For single-operator prediction, we compare against Roofline~\cite{williams2009roofline}, NeuSight~\cite{lee2025forecasting}\footnote{The original NeuSight is trained only on FP32 GEMMs; we retrain it with PipeWeave's FP16 dataset for a fair comparison.}, PipeWeave~\cite{pipeweave2026}, and GenZ~\cite{bambhaniya2024genz}.
NeuSight is trained on BF16/FP16 GEMM data from six PipeWeave GPUs, including A100.
The PipeWeave dataset covers A100 and Hopper-class machines, so PipeWeave is not a zero-shot baseline on those architectures.
For distributed kernels and end-to-end serving, we compare against PipeWeave and GenZ.
PipeWeave's collective model is a per-GPU random forest with no configurable $\alpha$--$\beta$ or topology parameters; among our targets, A100 and B6000 (RTX PRO 6000 Blackwell) have native PipeWeave collective datasets, while H200-NVL and B200 are unsupported and we use its H800 dataset as the closest available substitute.
For end-to-end serving, we provide PipeWeave with the required vendor hardware specifications.

\subsection{Single-Operator Prediction Accuracy}
\label{sec:model-accuracy}

\begin{figure}[t]
    \centering
    \includegraphics[width=\columnwidth]{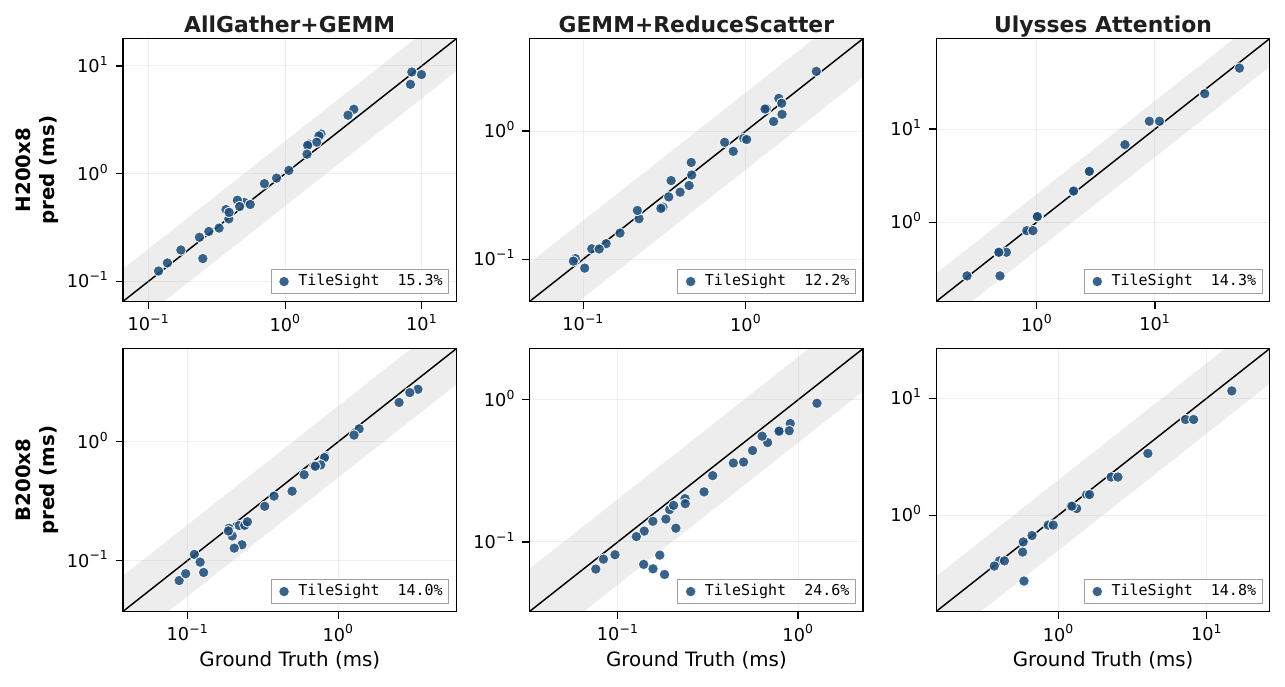}
    \caption{Fused compute-communication kernel prediction on H200$\times$8 and B200$\times$8 (AllGather+GEMM, GEMM+ReduceScatter, Ulysses Attention).}
    \label{fig:dist_kernel_compute}
\end{figure}

\begin{figure*}[t]
    \centering
    \includegraphics[width=0.98\textwidth]{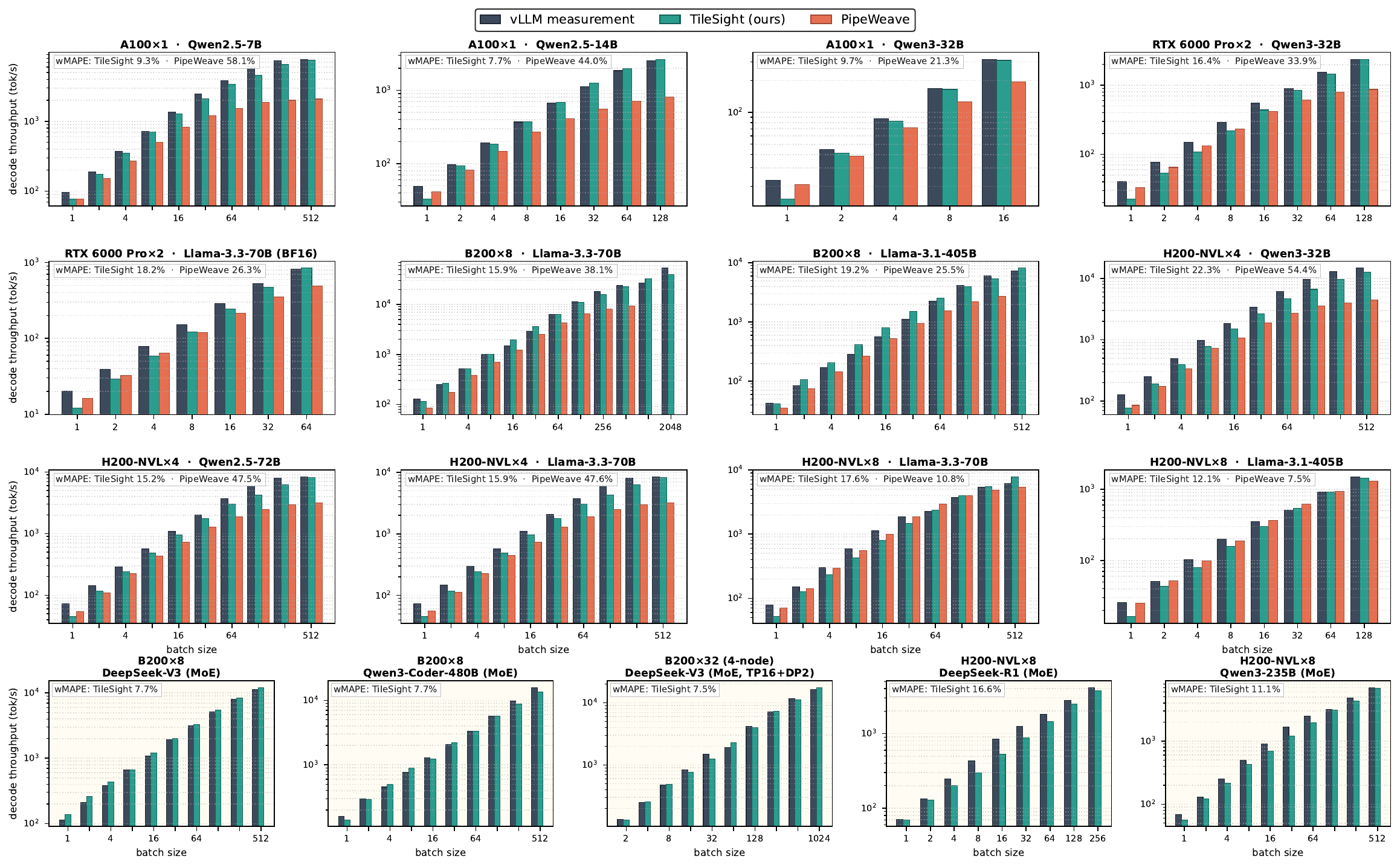}
    \caption{vLLM decode throughput prediction across dense LLMs, MoE models, and multi-node configurations. Dense rows cover A100$\times$1, B6000$\times$2, B200$\times$8, and H200-NVL. MoE rows cover B200$\times$8, B200$\times$32, and H200-NVL$\times$8. Bars compare measured vLLM tokens per second with \oursys{} and PipeWeave where supported. PipeWeave does not support MoE.}
    \label{fig:vllm_e2e_stack}
\end{figure*}

\begin{figure}[t]
    \centering
    \includegraphics[width=\linewidth]{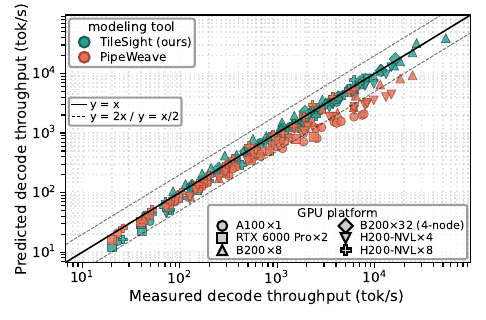}
    \caption{Predicted vs.\ measured decode throughput across all healthy configurations. \oursys{}: 13.52\% wMAPE overall. PipeWeave: 31.84\% wMAPE on supported dense rows.}
    \label{fig:vllm_e2e_scatter}
\end{figure}

Figure~\ref{fig:gemm_cross_scatter} evaluates 703 BF16/FP16 tensor-core GEMM shapes on A100, B200, B6000, and H200, using 
\texttt{cutlass\_profiler} measurements as ground truth after filtering stream-K and single-instruction, multiple-thread (SIMT) fallback paths.
\oursys{} achieves 12.35\% pooled MAPE, compared with 21.97\% for PipeWeave, 32.95\% for retrained NeuSight, 33.85\% for Roofline, and 34.89\% for GenZ.
\oursys{} is best on the newer B200, B6000, and H200 targets.
NeuSight narrowly leads on A100 because A100 appears in its training distribution, but this advantage does not transfer to newer GPUs, illustrating the overfitting risk of architecture-specific learned predictors.
On MI210, because CK provides no explicit rasterization (along-$M$/along-$N$) or swizzle control as \texttt{cutlass\_profiler} does, \oursys{} runs in its default cache mode, yet still leads at 23.4\% MAPE, ahead of PipeWeave (25.5\%), NeuSight (26.4\%), Roofline (38.8\%), and GenZ (40.4\%).
Non-GEMM fused operators are evaluated in the distributed and end-to-end workloads below.

Table~\ref{tab:fa3_final_comparison} compares \oursys{} with NCU on a fused FA-3 kernel.
The final model predicts latency within 2.7\% and tracks the major resource-utilization components, providing a compact sanity check for the tile-pipeline model on non-GEMM fused execution.

\begin{table*}[t]
\centering
\footnotesize
\caption{Performance improvements in \oursys{} diagnosed kernels}
\label{tab:triton_opt}
\begin{tabularx}{\textwidth}{
    >{\centering\arraybackslash}p{0.08\textwidth}
    >{\centering\arraybackslash}p{0.08\textwidth}
    >{\centering\arraybackslash}p{0.08\textwidth}
    >{\centering\arraybackslash}p{0.06\textwidth}
    >{\centering\arraybackslash}p{0.20\textwidth}
    >{\centering\arraybackslash}p{0.223\textwidth}
    >{\centering\arraybackslash}p{0.07\textwidth}
    >{\centering\arraybackslash}p{0.07\textwidth}
}
\toprule
\textbf{Kernel} & \textbf{Framework} & \textbf{Device} & \textbf{Baseline} & \textbf{Issue} & \textbf{Solution} & \textbf{Optimized} & \textbf{Speedup} \\
\midrule
ReLU     & Triton    & MI210  & 1.40ms & Indirect addr.  & Unroll addr.   & 1.10ms & 1.27× \\
Avg\_Pool & Triton    & MI210  & 0.20ms & Indirect addr. + Not Overlapped   & Unroll addr. + Small tile & 0.10ms & 2.00× \\
Avg\_Pool & Torch & MI210 & 0.15ms & Not Overlapped  & Small tile & 0.10ms & 1.50× \\
GEMM(M128) & CK & MI210  & 3.68ms & Not Overlapped  & Multi Thread Block per SM & 2.68ms & 1.37× \\
GEMM(K57344) & CK & MI210  & 55.63ms & Large K with L2 hit rate issue  & large tilek->1 TB per CU  & 51.90ms & 1.07× \\
RMS\_Norm & Torch.Compile & H100  & 0.21ms & Not Overlapped  & Multi Thread Block per SM  & 0.18ms & 1.17× \\
MLA(kv8192 b128 h128) & Triton & MI210 & 66.38ms & Tiling, Memory Alloc., SMEM Conflict   & Register alloc., larger Tile, Conflict Elim.  & 7.40ms & \textbf{8.97×} \\
\bottomrule
\end{tabularx}
\end{table*}

\begin{figure}[t]
    \centering
    \includegraphics[width=\columnwidth]{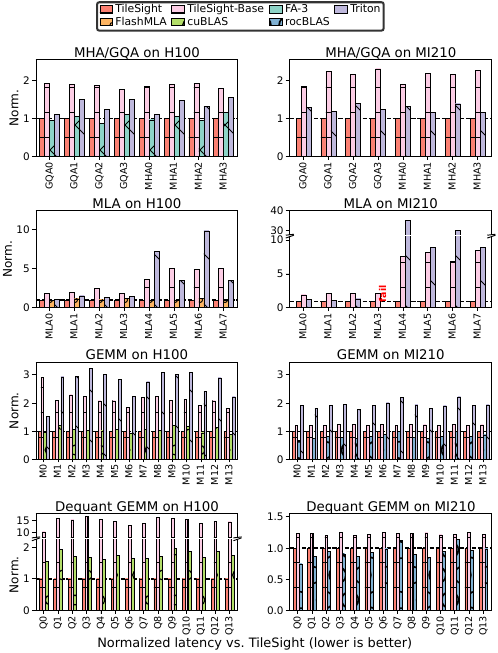}
    \caption{Kernel performance on H100 and MI210 when \oursys{} guides tile configuration selection in Triton and TileLang, replacing exhaustive autotuning. Reference lines are FlashAttention-3 for multi-head attention/grouped-query attention (MHA/GQA), FlashMLA for MLA, cuBLAS/rocBLAS for matrix multiplication, and vendor libraries for dequantized matrix multiplication.}
    \label{fig:microbenchmark_cafe}
\end{figure}

\begin{figure}[t]
    \centering
    \includegraphics[width=0.8\columnwidth]{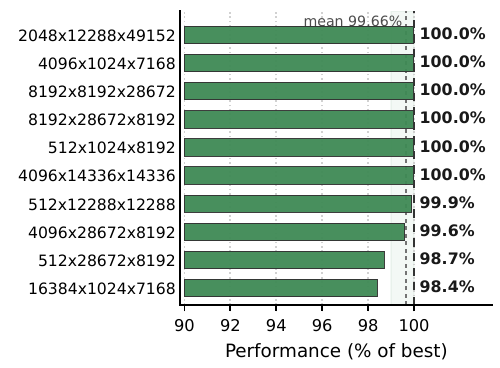}
    \caption{\oursys{} as cost model in TileLang: pruning 95\% of candidate schedules and retaining the predicted top 5\% reaches 99.66\% of exhaustive-search best performance on average across 10 LLaMA-derived GEMM-FP16 workloads.}
    \label{fig:gemm_cost_model}
\end{figure}

\subsection{L2 Cache Prediction Accuracy}
\label{ssec:l2_deep_dive}

Figure~\ref{fig:cache_hit_eval} evaluates tile reuse-distance cache modeling against NCU on 4{,}680 GEMM persistent-kernel cases.
With effective cache capacity calibrated by the bandwidth sweep in Figure~\ref{fig:cache_bw_sweep}, mean absolute L2 hit-rate error stays near one percentage point on every GPU: 1.46\,pp on A100, 0.88\,pp on H200, 1.05\,pp on B200, and 0.78\,pp on B6000 .
The results demonstrate the effectiveness of tile reuse-distance cache modeling.

\para{Effect of inter-SM execution skew.}
The reuse-distance model assumes tiles advance at a uniform rate, but SMs desynchronize and work on different $K$-slices at once, spreading the concurrently accessed tiles beyond L2; for deep-$K$ GEMMs this pushes the measured hit rate below \oursys{}'s lockstep prediction (e.g., a GEMM with $M{=}N{=}8192$, $K{=}28672$ on H200: 82\% predicted vs.\ 43\% measured).
Such configurations are rare, so aggregate error stays near one percentage point, but the model is systematically optimistic in this regime; we revisit it in \S\ref{sec:limitations}.

\subsection{Distributed Validation}
\label{ssec:dist_eval}

Figures~\ref{fig:dist_kernel_collective} and~\ref{fig:dist_kernel_compute} validate 304 distributed cases on H200$\times$8 and B200$\times$8: 152 pure collectives and 152 fused compute-communication kernels.
\oursys{} extracts logical source--destination exchanges, routes them over calibrated NVLink topologies, and evaluates each stage with the \(\alpha\)-\(\beta\) model from \S\ref{ssec:distributed}.
On pure collectives, \oursys{} achieves 12.22\% wMAPE, compared with 20.82\% for GenZ and 65.72\% for PipeWeave on supported rows.
PipeWeave has no native configurable H200/B200 backend for these collectives and falls back to an H800 random-forest model, so it cannot reflect the NVLink4/5 bandwidth differences in our machines.
For B200 Ulysses Attention, the local compute stage uses the source-aligned SM100 $128\!\times\!128$ FA4 tile pipeline with TMEM traffic, packed grids, and the sectioned LPT mapping, composed with four all-to-all stages.
On fused kernels, where both baselines are unsupported, \oursys{} achieves 14.83\% wMAPE.

\subsection{vLLM End-to-End Decode}
\label{ssec:e2e_cases}

Figures~\ref{fig:vllm_e2e_stack} and~\ref{fig:vllm_e2e_scatter} evaluate end-to-end vLLM decode throughput on 166 healthy configurations spanning dense, MoE, single-node, and multi-node serving.
The evaluated systems range from A100$\times$1 and B6000$\times$2 to B200$\times$32 and H200-NVL$\times$8, exercising both local tile execution and routed distributed stages.
Overall, \oursys{} achieves 13.52\% wMAPE, while PipeWeave with the B200 extension reaches 31.84\% wMAPE on 114/117 dense configurations. PipeWeave does not support MoE.
PipeWeave uses native collective datasets for A100 and B6000, but falls back to H800 for H200-NVL and B200.
For B200, we extend PipeWeave by supplying B200 hardware specifications while using its closest available H800 samples for GEMM-configuration lookup and its Hopper calculator.
The B200 extension produces valid predictions for 19/22 dense configurations. In the remaining three large-batch cases, the prefill RMSNorm sequence lengths exceed PipeWeave's 131K-token MLP training maximum. Although PipeWeave bounds its learned utilization factor to $[0,1]$ with a sigmoid, these out-of-range inputs drive it to zero, triggering division by zero and preventing robust end-to-end prediction for these cases. This highlights a robustness limitation of ML-based predictors when extrapolating to unseen cases.
\oursys{} achieves 7.5--18.0\% per-machine wMAPE and 10.35\% wMAPE on MoE configurations.

\subsection{Key Applications: Diagnosis and Cost Model}
\label{ssec:dsl_integration}

Due to its interpretable nature, \oursys{} can be used as a white-box optimization aid.
Figure~\ref{fig:microbenchmark_cafe} shows that tile configurations selected by \oursys{} can match or exceed strong vendor and expert baselines across attention, MLA, GEMM, and dequantized matmul kernels on H100 and MI210.
Figure~\ref{fig:gemm_cost_model} shows the same model used as a TileLang cost model: retaining the predicted top 5\% schedules prunes 95\% of candidates while reaching 99.66\% of exhaustive-search best performance on average.
This is especially useful on less-supported targets, where learned or vendor-tuned cost models provide weak guidance but the analytical model can still surface high-quality schedule candidates.

The diagnosis cases fall into four recurring bottleneck classes: indirect addressing, insufficient pipeline overlap, poor L2 locality, and architecture-specific memory-layout issues.
In each case, \oursys{} maps the bottleneck to concrete tile-level changes, such as address unrolling, tile-size adjustment, higher resident-block occupancy, or shared-memory/register-layout fixes.
Table~\ref{tab:triton_opt} summarizes diagnosis cases where \oursys{} identifies indirect addressing, pipeline stalls, and L2 locality bottlenecks, leading to 1.07--8.97$\times$ improvements.

\section{Related Work} \label{sec:related_work}

\paragraph{Tile-Centric Programming Frameworks.}
Triton~\cite{tillet2019triton}, TileLang~\cite{tilelang}, TileLink~\cite{zheng2025tilelink}, CUTLASS/CUTE~\cite{nvidia2024cutlass}, CuteDSL~\cite{cutlass}, ThunderKittens~\cite{spector2024thunderkittens}, FractalTensor~\cite{liu2024fractal_tensor}, and NVIDIA's CUDA Tile~\cite{cudatile} have driven GPU programming toward tile-centric abstractions.
Yet none ships a tile-centric performance model: Triton relies on black-box autotuning, TileLang on heuristics, and tritonBLAS~\cite{swann2025tritonblas} on GEMM-specific analytical selection.
\oursys{} fills this gap as a unified tile-centric cost model and diagnosis backend for these frameworks.

\paragraph{Performance Modeling and Prediction.}
Roofline~\cite{williams2009roofline} and its variants (e.g., GenZ~\cite{bambhaniya2024genz}, \cite{morgado2024carm, yuan2024llm, patwari2025forecasting, davies2025liminal}) provide useful first-order bounds for LLM inference but cannot distinguish kernels with different schedules at identical FLOP/byte counts, nor capture schedule-dependent effects such as L2 reuse under different tile orders.
Karami et al.~\cite{karami2025understanding} further show non-GEMM ops account for up to 74\% of inference latency, challenging GEMM-centric assumptions.
Dataflow exploration frameworks~\cite{parashar2019timeloop, gao2019tangram, kwon2020maestro, zheng2023tileflow, wu2022sparseloop, cai2023inter} model loop nests and data reuse for spatial accelerators but rely on simplified hardware assumptions that limit GPU applicability.
Hybrid and ML-based approaches---PipeWeave~\cite{pipeweave2026}, NeuSight~\cite{lee2025forecasting}, CDMPP~\cite{hu2024cdmpp}, TAO~\cite{pandey2024tao}, Omniwise~\cite{wang2025omniwise}, among others~\cite{geoffrey2021habitat, li2023path}---predict runtime via learned models (either end-to-end or as residuals on top of analytical estimates), often accurately but as black boxes without exposing \emph{why} a kernel is slow.
\oursys{} differs by being fully first-principles with no learned component, yet matches or exceeds these predictors' accuracy while offering schedule-aware, tile-granular diagnosis that decomposes performance into actionable components.

\paragraph{GPU Profiling and Instrumentation.}
Vendor profilers (Nsight Compute~\cite{nvidia_nsight_compute}, OmniPerf~\cite{amd_omniperf}) report metrics but little root-cause guidance.
KPerfIR~\cite{guan2025kperfir} and Neutrino~\cite{huang2025neutrino} advance compiler- and probe-based GPU instrumentation, while binary-level tools~\cite{shen2018cudaadvisor, zhou2020gpa_cgo, zhou2021gpa, zeni2024starlight} provide low-level visibility.
All are \emph{post-hoc}, requiring execution and unable to predict unseen configurations.
\oursys{} predicts performance before execution and maps bottlenecks to tile-level scheduling decisions.

\paragraph{Distributed Multi-GPU Performance Modeling.}
Vidur \cite{agrawal2024vidur}, Lumos~\cite{Liang2025LumosEP}, SimAI~\cite{Wang2025SimAIUA}, TokenSim~\cite{Wu2025TokenSimEH}, Maya~\cite{yarlagadda2025maya}, and Echo~\cite{Feng2024EchoSD} simulate distributed training or inference at scale using profiling-based kernel estimators,
while DistServe~\cite{zhong2024distserve}, 
CrossPipe\allowbreak \cite{Chen2025CrossPipeTO}, 
Sailor~\cite{strati2025sailor}, Metis~\cite{um2024metis}, and RAPID-LLM~\cite{Karfakis2025RAPIDLLMRP} optimize parallel strategies with various communication and scheduling models.
All treat single-GPU kernel execution as a black box.
\oursys{} operates at the complementary intra-kernel level, providing white-box tile-level cost estimation that can plug into these distributed simulators, while its own distributed extension composes tile-level predictions with communication models under a unified tile abstraction.

\section{Limitations and Future Work}
\label{sec:limitations}

\oursys{} targets regular, tile-structured programs whose runtime is dominated by resource utilization.
It does not model data-dependent control flow, highly irregular memory access, instruction-level compiler decisions, undocumented warp/cooperative thread array (CTA) scheduling, or closed-source runtime behavior.
The hardware abstraction focuses on throughput.
Latency-bound cases such as small-batch decode attention, and multi-die effects such as B200 SM-to-HBM affinity, require finer-grained latency and topology parameters.
\oursys{} also assumes tiles execute at a uniform rate across SMs; in reality SMs desynchronize, which makes L2 hit-rate prediction mildly optimistic for large-$K$ GEMMs (\S\ref{ssec:l2_deep_dive}).
Finally, although \oursys{} has been validated as a TileLang cost model on selected GEMM workloads, broader compiler integration and non-GEMM schedule search remain future work.

\section{Conclusion}
\label{sec:conclusion}

\oursys{} shows that the tile, already a universal GPU programming unit across Triton, TileLang, CUDA Tile, and CuteDSL, can also unify performance reasoning. With tile-level modeling of resource use, dependencies, cache reuse, and cross-device placement, \oursys{} accurately predicts performance from single kernels to multi-node clusters without per-architecture training or profiling.
The broader lesson is a first-principles one: begin with a compact set of physically grounded mechanisms, and let their composition explain complex execution. For regular tile-structured workloads, this approach can yield accurate and interpretable predictions that transfer across architectures.

\clearpage
\newpage
\bibliographystyle{ACM-Reference-Format}
\bibliography{refs}

\begin{thebibliography}{59}

\ifx \showCODEN    \undefined \def \showCODEN     #1{\unskip}     \fi
\ifx \showISBNx    \undefined \def \showISBNx     #1{\unskip}     \fi
\ifx \showISBNxiii \undefined \def \showISBNxiii  #1{\unskip}     \fi
\ifx \showISSN     \undefined \def \showISSN      #1{\unskip}     \fi
\ifx \showLCCN     \undefined \def \showLCCN      #1{\unskip}     \fi
\ifx \shownote     \undefined \def \shownote      #1{#1}          \fi
\ifx \showarticletitle \undefined \def \showarticletitle #1{#1}   \fi
\ifx \showURL      \undefined \def \showURL       {\relax}        \fi
\providecommand\bibfield[2]{#2}
\providecommand\bibinfo[2]{#2}
\providecommand\natexlab[1]{#1}
\providecommand\showeprint[2][]{arXiv:#2}

\bibitem[Abramowitz and Stegun(1965)]%
        {abramowitz1965handbook}
\bibfield{author}{\bibinfo{person}{Milton Abramowitz} {and}
  \bibinfo{person}{Irene~A Stegun}.} \bibinfo{year}{1965}\natexlab{}.
\newblock \bibinfo{booktitle}{\emph{Handbook of mathematical functions: with
  formulas, graphs, and mathematical tables}}. Vol.~\bibinfo{volume}{55}.
\newblock \bibinfo{publisher}{Courier Corporation}.
\newblock


\bibitem[{Advanced Micro Devices, Inc.}(2025)]%
        {amd_omniperf}
\bibfield{author}{\bibinfo{person}{{Advanced Micro Devices, Inc.}}}
  \bibinfo{year}{2025}\natexlab{}.
\newblock \bibinfo{title}{Omniperf Documentation}.
\newblock
\urldef\tempurl%
\url{https://rocm.docs.amd.com/projects/omniperf/en/docs-6.2.1/what-is-omniperf.html}
\showURL{%
\tempurl}
\newblock
\shownote{Accessed: 2025-04-15}.


\bibitem[Agrawal et~al\mbox{.}(2024)]%
        {agrawal2024vidur}
\bibfield{author}{\bibinfo{person}{Amey Agrawal}, \bibinfo{person}{Nitin
  Kedia}, \bibinfo{person}{Jayashree Mohan}, \bibinfo{person}{Ashish Panwar},
  \bibinfo{person}{Nipun Kwatra}, \bibinfo{person}{Bhargav~S Gulavani},
  \bibinfo{person}{Ramachandran Ramjee}, {and} \bibinfo{person}{Alexey
  Tumanov}.} \bibinfo{year}{2024}\natexlab{}.
\newblock \showarticletitle{Vidur: A large-scale simulation framework for llm
  inference}.
\newblock \bibinfo{journal}{\emph{Proceedings of Machine Learning and Systems}}
   \bibinfo{volume}{6} (\bibinfo{year}{2024}), \bibinfo{pages}{351--366}.
\newblock


\bibitem[Arafa et~al\mbox{.}(2020)]%
        {arafa2020fast}
\bibfield{author}{\bibinfo{person}{Yehia Arafa}, \bibinfo{person}{Abdel-Hameed
  Badawy}, \bibinfo{person}{Gopinath Chennupati}, \bibinfo{person}{Atanu
  Barai}, \bibinfo{person}{Nandakishore Santhi}, {and} \bibinfo{person}{Stephan
  Eidenbenz}.} \bibinfo{year}{2020}\natexlab{}.
\newblock \showarticletitle{Fast, accurate, and scalable memory modeling of
  GPGPUs using reuse profiles}. In \bibinfo{booktitle}{\emph{Proceedings of the
  34th ACM International Conference on supercomputing}}.
  \bibinfo{pages}{1--12}.
\newblock


\bibitem[Arafa et~al\mbox{.}(2019)]%
        {arafa2019gpus}
\bibfield{author}{\bibinfo{person}{Yehia Arafa}, \bibinfo{person}{Gopinath
  Chennupati}, \bibinfo{person}{Atanu Barai}, \bibinfo{person}{Abdel-Hameed~A
  Badawy}, \bibinfo{person}{Nandakishore Santhi}, {and}
  \bibinfo{person}{Stephan Eidenbenz}.} \bibinfo{year}{2019}\natexlab{}.
\newblock \showarticletitle{Gpus cache performance estimation using reuse
  distance analysis}. In \bibinfo{booktitle}{\emph{2019 IEEE 38th International
  Performance Computing and Communications Conference (IPCCC)}}. IEEE,
  \bibinfo{pages}{1--8}.
\newblock


\bibitem[Bambhaniya et~al\mbox{.}(2024)]%
        {bambhaniya2024genz}
\bibfield{author}{\bibinfo{person}{Abhimanyu Bambhaniya},
  \bibinfo{person}{Ritik Raj}, \bibinfo{person}{Geonhwa Jeong},
  \bibinfo{person}{Souvik Kundu}, \bibinfo{person}{Sudarshan Srinivasan},
  \bibinfo{person}{Suvinay Subramanian}, \bibinfo{person}{Midhilesh
  Elavazhagan}, \bibinfo{person}{Madhu Kumar}, {and} \bibinfo{person}{Tushar
  Krishna}.} \bibinfo{year}{2024}\natexlab{}.
\newblock \showarticletitle{Demystifying AI Platform Design for Distributed
  Inference of Next-Generation LLM models}.
\newblock \bibinfo{journal}{\emph{arXiv preprint arXiv:2406.01698}}
  (\bibinfo{year}{2024}).
\newblock


\bibitem[Cai et~al\mbox{.}(2023)]%
        {cai2023inter}
\bibfield{author}{\bibinfo{person}{Jingwei Cai}, \bibinfo{person}{Yuchen Wei},
  \bibinfo{person}{Zuotong Wu}, \bibinfo{person}{Sen Peng}, {and}
  \bibinfo{person}{Kaisheng Ma}.} \bibinfo{year}{2023}\natexlab{}.
\newblock \showarticletitle{Inter-layer scheduling space definition and
  exploration for tiled accelerators}. In \bibinfo{booktitle}{\emph{Proceedings
  of the 50th Annual International Symposium on Computer Architecture}}.
  \bibinfo{pages}{1--17}.
\newblock


\bibitem[Chen et~al\mbox{.}(2025)]%
        {Chen2025CrossPipeTO}
\bibfield{author}{\bibinfo{person}{Tiancheng Chen}, \bibinfo{person}{Ale{\v{s}}
  Kub{\'i}{\v{c}}ek}, \bibinfo{person}{Langwen Huang}, {and}
  \bibinfo{person}{Torsten Hoefler}.} \bibinfo{year}{2025}\natexlab{}.
\newblock \showarticletitle{CrossPipe: Towards Optimal Pipeline Schedules for
  Cross-Datacenter Training}. In \bibinfo{booktitle}{\emph{USENIX Annual
  Technical Conference}}.
\newblock
\urldef\tempurl%
\url{https://api.semanticscholar.org/CorpusID:280049543}
\showURL{%
\tempurl}


\bibitem[Conte et~al\mbox{.}(1998)]%
        {conte1998combining}
\bibfield{author}{\bibinfo{person}{Thomas~M. Conte}, \bibinfo{person}{Mary~Ann
  Hirsch}, {and} \bibinfo{person}{W-MW Hwu}.} \bibinfo{year}{1998}\natexlab{}.
\newblock \showarticletitle{Combining trace sampling with single pass methods
  for efficient cache simulation}.
\newblock \bibinfo{journal}{\emph{IEEE Trans. Comput.}} \bibinfo{volume}{47},
  \bibinfo{number}{6} (\bibinfo{year}{1998}), \bibinfo{pages}{714--720}.
\newblock


\bibitem[Contributors(2025)]%
        {tilelang}
\bibfield{author}{\bibinfo{person}{Tile-AI Contributors}.}
  \bibinfo{year}{2025}\natexlab{}.
\newblock \bibinfo{title}{TileLang: A Domain-Specific Language for
  High-Performance GPU/CPU Kernels}.
\newblock \bibinfo{howpublished}{\url{https://github.com/tile-ai/tilelang}}.
\newblock


\bibitem[Corporation(2024)]%
        {nvidia2024cutlass}
\bibfield{author}{\bibinfo{person}{NVIDIA Corporation}.}
  \bibinfo{year}{2024}\natexlab{}.
\newblock \bibinfo{title}{CUTLASS: CUDA Templates for Linear Algebra
  Subroutines}.
\newblock \bibinfo{howpublished}{\url{https://github.com/NVIDIA/cutlass}}.
\newblock


\bibitem[CUTLASS(acce)]%
        {cutlass}
CUTLASS \bibinfo{year}{acce}\natexlab{}.
\newblock \bibinfo{title}{{NVIDIA CUTLASS}}.
\newblock
\newblock
\shownote{\url{https://github.com/NVIDIA/cutlass}}.


\bibitem[Davies et~al\mbox{.}(2025)]%
        {davies2025liminal}
\bibfield{author}{\bibinfo{person}{Michael Davies}, \bibinfo{person}{Neal
  Crago}, \bibinfo{person}{Karthikeyan Sankaralingam}, {and}
  \bibinfo{person}{Christos Kozyrakis}.} \bibinfo{year}{2025}\natexlab{}.
\newblock \showarticletitle{LIMINAL: Exploring The Frontiers of LLM Decode
  Performance}.
\newblock \bibinfo{journal}{\emph{arXiv preprint arXiv:2507.14397}}
  (\bibinfo{year}{2025}).
\newblock


\bibitem[Feng et~al\mbox{.}(2024)]%
        {Feng2024EchoSD}
\bibfield{author}{\bibinfo{person}{Yicheng Feng}, \bibinfo{person}{Yuetao
  Chen}, \bibinfo{person}{Kaiwen Chen}, \bibinfo{person}{Jingzong Li},
  \bibinfo{person}{Tianyuan Wu}, \bibinfo{person}{Peng Cheng},
  \bibinfo{person}{Chuan Wu}, \bibinfo{person}{Wei Wang},
  \bibinfo{person}{Tsung-Yi Ho}, {and} \bibinfo{person}{Hong Xu}.}
  \bibinfo{year}{2024}\natexlab{}.
\newblock \showarticletitle{Echo: Simulating Distributed Training At Scale}.
\newblock \bibinfo{journal}{\emph{ArXiv}}  \bibinfo{volume}{abs/2412.12487}
  (\bibinfo{year}{2024}).
\newblock
\urldef\tempurl%
\url{https://api.semanticscholar.org/CorpusID:274789139}
\showURL{%
\tempurl}


\bibitem[{Futu News}(2026)]%
        {cudatile_keynote}
\bibfield{author}{\bibinfo{person}{{Futu News}}.}
  \bibinfo{year}{2026}\natexlab{}.
\newblock \bibinfo{title}{{NVIDIA launches CUDA 13.1 and CUDA Tile; Jensen
  Huang calls it the most significant advancement since CUDA's introduction
  $\sim$20 years ago}}.
\newblock \bibinfo{howpublished}{Online news brief}.
\newblock
\urldef\tempurl%
\url{https://news.futunn.com/en/post/65885271/futu-morning-brief-the-most-critical-week-of-the-year}
\showURL{%
\tempurl}
\newblock
\shownote{Accessed: 2026-05-19}.


\bibitem[Gao et~al\mbox{.}(2019)]%
        {gao2019tangram}
\bibfield{author}{\bibinfo{person}{Mingyu Gao}, \bibinfo{person}{Xuan Yang},
  \bibinfo{person}{Jing Pu}, \bibinfo{person}{Mark Horowitz}, {and}
  \bibinfo{person}{Christos Kozyrakis}.} \bibinfo{year}{2019}\natexlab{}.
\newblock \showarticletitle{Tangram: Optimized coarse-grained dataflow for
  scalable nn accelerators}. In \bibinfo{booktitle}{\emph{Proceedings of the
  Twenty-Fourth International Conference on Architectural Support for
  Programming Languages and Operating Systems}}. \bibinfo{pages}{807--820}.
\newblock


\bibitem[Geoffrey et~al\mbox{.}(2021)]%
        {geoffrey2021habitat}
\bibfield{author}{\bibinfo{person}{X~Yu Geoffrey}, \bibinfo{person}{Yubo Gao},
  \bibinfo{person}{Pavel Golikov}, {and} \bibinfo{person}{Gennady Pekhimenko}.}
  \bibinfo{year}{2021}\natexlab{}.
\newblock \showarticletitle{Habitat: A $\{$Runtime-Based$\}$ computational
  performance predictor for deep neural network training}. In
  \bibinfo{booktitle}{\emph{2021 USENIX Annual Technical Conference (USENIX ATC
  21)}}. \bibinfo{pages}{503--521}.
\newblock


\bibitem[Guan et~al\mbox{.}(2025)]%
        {guan2025kperfir}
\bibfield{author}{\bibinfo{person}{Yue Guan}, \bibinfo{person}{Yuanwei Fang},
  \bibinfo{person}{Keren Zhou}, \bibinfo{person}{Corbin Robeck},
  \bibinfo{person}{Manman Ren}, \bibinfo{person}{Zhongkai Yu},
  \bibinfo{person}{Yufei Ding}, {and} \bibinfo{person}{Adnan Aziz}.}
  \bibinfo{year}{2025}\natexlab{}.
\newblock \showarticletitle{KPerfIR: Towards an Open and Compiler-centric
  Ecosystem for GPU Kernel Performance Tooling on Modern AI Workloads}.
\newblock \bibinfo{journal}{\emph{arXiv preprint arXiv:2505.21661}}
  (\bibinfo{year}{2025}).
\newblock


\bibitem[Hu et~al\mbox{.}(2024)]%
        {hu2024cdmpp}
\bibfield{author}{\bibinfo{person}{Hanpeng Hu}, \bibinfo{person}{Junwei Su},
  \bibinfo{person}{Juntao Zhao}, \bibinfo{person}{Yanghua Peng},
  \bibinfo{person}{Yibo Zhu}, \bibinfo{person}{Haibin Lin}, {and}
  \bibinfo{person}{Chuan Wu}.} \bibinfo{year}{2024}\natexlab{}.
\newblock \showarticletitle{CDMPP: A device-model agnostic framework for
  latency prediction of tensor programs}. In
  \bibinfo{booktitle}{\emph{Proceedings of the Nineteenth European Conference
  on Computer Systems}}. \bibinfo{pages}{1054--1074}.
\newblock


\bibitem[Huang and Wu(2025)]%
        {huang2025neutrino}
\bibfield{author}{\bibinfo{person}{Songlin Huang} {and}
  \bibinfo{person}{Chenshu Wu}.} \bibinfo{year}{2025}\natexlab{}.
\newblock \showarticletitle{Neutrino: Fine-grained $\{$GPU$\}$ Kernel Profiling
  via Programmable Probing}. In \bibinfo{booktitle}{\emph{19th USENIX Symposium
  on Operating Systems Design and Implementation (OSDI 25)}}.
  \bibinfo{pages}{331--355}.
\newblock


\bibitem[Karami et~al\mbox{.}(2025)]%
        {karami2025understanding}
\bibfield{author}{\bibinfo{person}{Rachid Karami}, \bibinfo{person}{Sheng-Chun
  Kao}, {and} \bibinfo{person}{Hyoukjun Kwon}.}
  \bibinfo{year}{2025}\natexlab{}.
\newblock \showarticletitle{Understanding the Performance Horizon of the Latest
  ML Workloads with NonGEMM Workloads}. In \bibinfo{booktitle}{\emph{2025 IEEE
  International Symposium on Performance Analysis of Systems and Software
  (ISPASS)}}. IEEE, \bibinfo{pages}{1--14}.
\newblock


\bibitem[Karfakis et~al\mbox{.}(2025)]%
        {Karfakis2025RAPIDLLMRP}
\bibfield{author}{\bibinfo{person}{George Karfakis}, \bibinfo{person}{Faraz
  Tahmasebi}, \bibinfo{person}{Bin Chen}, \bibinfo{person}{Lime Yao},
  \bibinfo{person}{Saptarshi Mitra}, \bibinfo{person}{Tian Pan},
  \bibinfo{person}{Hyoukjun Kwon}, {and} \bibinfo{person}{Puneet Gupta}.}
  \bibinfo{year}{2025}\natexlab{}.
\newblock \showarticletitle{RAPID-LLM: Resilience-Aware Performance analysis of
  Infrastructure for Distributed LLM Training and Inference}.
\newblock \bibinfo{journal}{\emph{ArXiv}}  \bibinfo{volume}{abs/2512.19606}
  (\bibinfo{year}{2025}).
\newblock
\urldef\tempurl%
\url{https://api.semanticscholar.org/CorpusID:284077588}
\showURL{%
\tempurl}


\bibitem[Kwon et~al\mbox{.}(2020)]%
        {kwon2020maestro}
\bibfield{author}{\bibinfo{person}{Hyoukjun Kwon}, \bibinfo{person}{Prasanth
  Chatarasi}, \bibinfo{person}{Vivek Sarkar}, \bibinfo{person}{Tushar Krishna},
  \bibinfo{person}{Michael Pellauer}, {and} \bibinfo{person}{Angshuman
  Parashar}.} \bibinfo{year}{2020}\natexlab{}.
\newblock \showarticletitle{Maestro: A data-centric approach to understand
  reuse, performance, and hardware cost of dnn mappings}.
\newblock \bibinfo{journal}{\emph{IEEE micro}} \bibinfo{volume}{40},
  \bibinfo{number}{3} (\bibinfo{year}{2020}), \bibinfo{pages}{20--29}.
\newblock


\bibitem[Lam et~al\mbox{.}(1991)]%
        {lam1991cache}
\bibfield{author}{\bibinfo{person}{Monica~D Lam}, \bibinfo{person}{Edward~E
  Rothberg}, {and} \bibinfo{person}{Michael~E Wolf}.}
  \bibinfo{year}{1991}\natexlab{}.
\newblock \showarticletitle{The cache performance and optimizations of blocked
  algorithms}.
\newblock \bibinfo{journal}{\emph{ACM SIGOPS Operating Systems Review}}
  \bibinfo{volume}{25}, \bibinfo{number}{Special Issue} (\bibinfo{year}{1991}),
  \bibinfo{pages}{63--74}.
\newblock


\bibitem[Lee et~al\mbox{.}(2025)]%
        {lee2025forecasting}
\bibfield{author}{\bibinfo{person}{Seonho Lee}, \bibinfo{person}{Amar
  Phanishayee}, {and} \bibinfo{person}{Divya Mahajan}.}
  \bibinfo{year}{2025}\natexlab{}.
\newblock \showarticletitle{Forecasting GPU Performance for Deep Learning
  Training and Inference}. In \bibinfo{booktitle}{\emph{Proceedings of the 30th
  ACM International Conference on Architectural Support for Programming
  Languages and Operating Systems, Volume 1}}. \bibinfo{pages}{493--508}.
\newblock


\bibitem[Li et~al\mbox{.}(2023)]%
        {li2023path}
\bibfield{author}{\bibinfo{person}{Ying Li}, \bibinfo{person}{Yifan Sun}, {and}
  \bibinfo{person}{Adwait Jog}.} \bibinfo{year}{2023}\natexlab{}.
\newblock \showarticletitle{Path Forward Beyond Simulators: Fast and Accurate
  GPU Execution Time Prediction for DNN Workloads}. In
  \bibinfo{booktitle}{\emph{Proceedings of the 56th Annual IEEE/ACM
  International Symposium on Microarchitecture}}. \bibinfo{pages}{380--394}.
\newblock


\bibitem[Liang et~al\mbox{.}(2025)]%
        {Liang2025LumosEP}
\bibfield{author}{\bibinfo{person}{Mingyu Liang}, \bibinfo{person}{Hiwot~Tadese
  Kassa}, \bibinfo{person}{Wenyin Fu}, \bibinfo{person}{Brian Coutinho},
  \bibinfo{person}{Louis Feng}, {and} \bibinfo{person}{Christina Delimitrou}.}
  \bibinfo{year}{2025}\natexlab{}.
\newblock \showarticletitle{Lumos: Efficient Performance Modeling and
  Estimation for Large-scale LLM Training}.
\newblock \bibinfo{journal}{\emph{ArXiv}}  \bibinfo{volume}{abs/2504.09307}
  (\bibinfo{year}{2025}).
\newblock
\urldef\tempurl%
\url{https://api.semanticscholar.org/CorpusID:277781663}
\showURL{%
\tempurl}


\bibitem[Liu et~al\mbox{.}(2024)]%
        {liu2024fractal_tensor}
\bibfield{author}{\bibinfo{person}{Siran Liu}, \bibinfo{person}{Chengxiang Qi},
  \bibinfo{person}{Ying Cao}, \bibinfo{person}{Chao Yang},
  \bibinfo{person}{Weifang Hu}, \bibinfo{person}{Xuanhua Shi},
  \bibinfo{person}{Fan Yang}, {and} \bibinfo{person}{Mao Yang}.}
  \bibinfo{year}{2024}\natexlab{}.
\newblock \showarticletitle{Uncovering nested data parallelism and data reuse
  in dnn computation with fractaltensor}. In
  \bibinfo{booktitle}{\emph{Proceedings of the ACM SIGOPS 30th Symposium on
  Operating Systems Principles}}. \bibinfo{pages}{160--177}.
\newblock


\bibitem[Morgado et~al\mbox{.}(2024)]%
        {morgado2024carm}
\bibfield{author}{\bibinfo{person}{Jos{\'e} Morgado}, \bibinfo{person}{Leonel
  Sousa}, {and} \bibinfo{person}{Aleksandar Ilic}.}
  \bibinfo{year}{2024}\natexlab{}.
\newblock \showarticletitle{CARM tool: cache-aware roofline model automatic
  benchmarking and application analysis}. In \bibinfo{booktitle}{\emph{2024
  IEEE International Symposium on Workload Characterization (IISWC)}}. IEEE,
  \bibinfo{pages}{68--81}.
\newblock


\bibitem[Niu et~al\mbox{.}(2012)]%
        {niu2012parda}
\bibfield{author}{\bibinfo{person}{Qingpeng Niu}, \bibinfo{person}{James
  Dinan}, \bibinfo{person}{Qingda Lu}, {and} \bibinfo{person}{Ponnuswamy
  Sadayappan}.} \bibinfo{year}{2012}\natexlab{}.
\newblock \showarticletitle{PARDA: A fast parallel reuse distance analysis
  algorithm}. In \bibinfo{booktitle}{\emph{2012 IEEE 26th International
  Parallel and Distributed Processing Symposium}}. IEEE,
  \bibinfo{pages}{1284--1294}.
\newblock


\bibitem[Nugteren et~al\mbox{.}(2014)]%
        {nugteren2014detailed}
\bibfield{author}{\bibinfo{person}{Cedric Nugteren}, \bibinfo{person}{Gert-Jan
  Van~den Braak}, \bibinfo{person}{Henk Corporaal}, {and}
  \bibinfo{person}{Henri Bal}.} \bibinfo{year}{2014}\natexlab{}.
\newblock \showarticletitle{A detailed GPU cache model based on reuse distance
  theory}. In \bibinfo{booktitle}{\emph{2014 IEEE 20th International Symposium
  on High Performance Computer Architecture (HPCA)}}. IEEE,
  \bibinfo{pages}{37--48}.
\newblock


\bibitem[{NVIDIA Corporation}(2025)]%
        {nvidia_nsight_compute}
\bibfield{author}{\bibinfo{person}{{NVIDIA Corporation}}.}
  \bibinfo{year}{2025}\natexlab{}.
\newblock \bibinfo{title}{NVIDIA Nsight Compute}.
\newblock
\urldef\tempurl%
\url{https://developer.nvidia.com/nsight-compute}
\showURL{%
\tempurl}
\newblock
\shownote{Accessed: 2025-04-15}.


\bibitem[{NVIDIA Corporation}(2026)]%
        {cudatile}
\bibfield{author}{\bibinfo{person}{{NVIDIA Corporation}}.}
  \bibinfo{year}{2026}\natexlab{}.
\newblock \bibinfo{title}{{CUDA Tile | NVIDIA Developer}}.
\newblock
\urldef\tempurl%
\url{https://developer.nvidia.com/cuda/tile}
\showURL{%
\tempurl}
\newblock
\shownote{Accessed: 2026-04-13}.


\bibitem[Pandey et~al\mbox{.}(2024)]%
        {pandey2024tao}
\bibfield{author}{\bibinfo{person}{Santosh Pandey}, \bibinfo{person}{Amir
  Yazdanbakhsh}, {and} \bibinfo{person}{Hang Liu}.}
  \bibinfo{year}{2024}\natexlab{}.
\newblock \showarticletitle{Tao: re-thinking DL-based microarchitecture
  simulation}.
\newblock \bibinfo{journal}{\emph{Proceedings of the ACM on Measurement and
  Analysis of Computing Systems}} \bibinfo{volume}{8}, \bibinfo{number}{2}
  (\bibinfo{year}{2024}), \bibinfo{pages}{1--25}.
\newblock


\bibitem[Parashar et~al\mbox{.}(2019)]%
        {parashar2019timeloop}
\bibfield{author}{\bibinfo{person}{Angshuman Parashar},
  \bibinfo{person}{Priyanka Raina}, \bibinfo{person}{Yakun~Sophia Shao},
  \bibinfo{person}{Yu-Hsin Chen}, \bibinfo{person}{Victor~A Ying},
  \bibinfo{person}{Anurag Mukkara}, \bibinfo{person}{Rangharajan Venkatesan},
  \bibinfo{person}{Brucek Khailany}, \bibinfo{person}{Stephen~W Keckler}, {and}
  \bibinfo{person}{Joel Emer}.} \bibinfo{year}{2019}\natexlab{}.
\newblock \showarticletitle{Timeloop: A systematic approach to dnn accelerator
  evaluation}. In \bibinfo{booktitle}{\emph{2019 IEEE international symposium
  on performance analysis of systems and software (ISPASS)}}. IEEE,
  \bibinfo{pages}{304--315}.
\newblock


\bibitem[Patwari et~al\mbox{.}(2025)]%
        {patwari2025forecasting}
\bibfield{author}{\bibinfo{person}{Rajeev Patwari}, \bibinfo{person}{Ashish
  Sirasao}, {and} \bibinfo{person}{Devleena Das}.}
  \bibinfo{year}{2025}\natexlab{}.
\newblock \showarticletitle{Forecasting LLM inference performance via
  hardware-agnostic analytical modeling}.
\newblock \bibinfo{journal}{\emph{arXiv preprint arXiv:2508.00904}}
  (\bibinfo{year}{2025}).
\newblock


\bibitem[Shen et~al\mbox{.}(2018)]%
        {shen2018cudaadvisor}
\bibfield{author}{\bibinfo{person}{Du Shen}, \bibinfo{person}{Shuaiwen~Leon
  Song}, \bibinfo{person}{Ang Li}, {and} \bibinfo{person}{Xu Liu}.}
  \bibinfo{year}{2018}\natexlab{}.
\newblock \showarticletitle{Cudaadvisor: Llvm-based runtime profiling for
  modern gpus}. In \bibinfo{booktitle}{\emph{Proceedings of the 2018
  International Symposium on Code Generation and Optimization}}.
  \bibinfo{pages}{214--227}.
\newblock


\bibitem[Spector et~al\mbox{.}(2024)]%
        {spector2024thunderkittens}
\bibfield{author}{\bibinfo{person}{Benjamin~F Spector}, \bibinfo{person}{Simran
  Arora}, \bibinfo{person}{Aaryan Singhal}, \bibinfo{person}{Daniel~Y Fu},
  {and} \bibinfo{person}{Christopher R{\'e}}.} \bibinfo{year}{2024}\natexlab{}.
\newblock \showarticletitle{ThunderKittens: Simple, Fast, and Adorable AI
  Kernels}.
\newblock \bibinfo{journal}{\emph{arXiv preprint arXiv:2410.20399}}
  (\bibinfo{year}{2024}).
\newblock


\bibitem[Strati et~al\mbox{.}(2025)]%
        {strati2025sailor}
\bibfield{author}{\bibinfo{person}{Foteini Strati}, \bibinfo{person}{Zhendong
  Zhang}, \bibinfo{person}{George Manos}, \bibinfo{person}{Ixeia~S{\'a}nchez
  P{\'e}riz}, \bibinfo{person}{Qinghao Hu}, \bibinfo{person}{Tiancheng Chen},
  \bibinfo{person}{Berk Buzcu}, \bibinfo{person}{Song Han},
  \bibinfo{person}{Pamela Delgado}, {and} \bibinfo{person}{Ana Klimovic}.}
  \bibinfo{year}{2025}\natexlab{}.
\newblock \showarticletitle{Sailor: Automating distributed training over
  dynamic, heterogeneous, and geo-distributed clusters}. In
  \bibinfo{booktitle}{\emph{Proceedings of the ACM SIGOPS 31st Symposium on
  Operating Systems Principles}}. \bibinfo{pages}{204--220}.
\newblock


\bibitem[Sul et~al\mbox{.}(2025)]%
        {sul2025parallelkittens}
\bibfield{author}{\bibinfo{person}{Stuart~H Sul}, \bibinfo{person}{Simran
  Arora}, \bibinfo{person}{Benjamin~F Spector}, {and}
  \bibinfo{person}{Christopher R{\'e}}.} \bibinfo{year}{2025}\natexlab{}.
\newblock \showarticletitle{ParallelKittens: Systematic and Practical
  Simplification of Multi-GPU AI Kernels}.
\newblock \bibinfo{journal}{\emph{arXiv preprint arXiv:2511.13940}}
  (\bibinfo{year}{2025}).
\newblock


\bibitem[Svedas et~al\mbox{.}(2025)]%
        {svedas2025survey}
\bibfield{author}{\bibinfo{person}{Jonas Svedas}, \bibinfo{person}{Hannah
  Watson}, \bibinfo{person}{Nathan Laubeuf}, \bibinfo{person}{Diksha
  Moolchandani}, \bibinfo{person}{Abubakr Nada}, \bibinfo{person}{Arjun Singh},
  \bibinfo{person}{Dwaipayan Biswas}, \bibinfo{person}{James Myers}, {and}
  \bibinfo{person}{Debjyoti Bhattacharjee}.} \bibinfo{year}{2025}\natexlab{}.
\newblock \showarticletitle{A survey of end-to-end modeling for distributed DNN
  training: Workloads, simulators, and TCO}.
\newblock \bibinfo{journal}{\emph{arXiv preprint arXiv:2506.09275}}
  (\bibinfo{year}{2025}).
\newblock


\bibitem[Swann et~al\mbox{.}(2025)]%
        {swann2025tritonblas}
\bibfield{author}{\bibinfo{person}{Ryan Swann}, \bibinfo{person}{Muhammad
  Osama}, \bibinfo{person}{Xiaohu Guo}, \bibinfo{person}{Bryant Nelson},
  \bibinfo{person}{Lixun Zhang}, \bibinfo{person}{Alex Brown},
  \bibinfo{person}{Yen Ong}, \bibinfo{person}{Ali Yazdani},
  \bibinfo{person}{Sean Siddens}, \bibinfo{person}{Ganesh Dasika},
  {et~al\mbox{.}}} \bibinfo{year}{2025}\natexlab{}.
\newblock \showarticletitle{tritonBLAS: Triton-based Analytical Approach for
  GEMM Kernel Parameter Selection}.
\newblock \bibinfo{journal}{\emph{arXiv preprint arXiv:2512.04226}}
  (\bibinfo{year}{2025}).
\newblock


\bibitem[Thakur et~al\mbox{.}(2005)]%
        {thakur2005optimization}
\bibfield{author}{\bibinfo{person}{Rajeev Thakur}, \bibinfo{person}{Rolf
  Rabenseifner}, {and} \bibinfo{person}{William Gropp}.}
  \bibinfo{year}{2005}\natexlab{}.
\newblock \showarticletitle{Optimization of collective communication operations
  in MPICH}.
\newblock \bibinfo{journal}{\emph{The International Journal of High Performance
  Computing Applications}} \bibinfo{volume}{19}, \bibinfo{number}{1}
  (\bibinfo{year}{2005}), \bibinfo{pages}{49--66}.
\newblock


\bibitem[Tillet et~al\mbox{.}(2019)]%
        {tillet2019triton}
\bibfield{author}{\bibinfo{person}{Philippe Tillet},
  \bibinfo{person}{Hsiang-Tsung Kung}, {and} \bibinfo{person}{David Cox}.}
  \bibinfo{year}{2019}\natexlab{}.
\newblock \showarticletitle{Triton: an intermediate language and compiler for
  tiled neural network computations}. In \bibinfo{booktitle}{\emph{Proceedings
  of the 3rd ACM SIGPLAN International Workshop on Machine Learning and
  Programming Languages}}. \bibinfo{pages}{10--19}.
\newblock


\bibitem[Um et~al\mbox{.}(2024)]%
        {um2024metis}
\bibfield{author}{\bibinfo{person}{Taegeon Um}, \bibinfo{person}{Byungsoo Oh},
  \bibinfo{person}{Minyoung Kang}, \bibinfo{person}{Woo-Yeon Lee},
  \bibinfo{person}{Goeun Kim}, \bibinfo{person}{Dongseob Kim},
  \bibinfo{person}{Youngtaek Kim}, \bibinfo{person}{Mohd Muzzammil}, {and}
  \bibinfo{person}{Myeongjae Jeon}.} \bibinfo{year}{2024}\natexlab{}.
\newblock \showarticletitle{Metis: Fast automatic distributed training on
  heterogeneous $\{$GPUs$\}$}. In \bibinfo{booktitle}{\emph{2024 USENIX Annual
  Technical Conference (USENIX ATC 24)}}. \bibinfo{pages}{563--578}.
\newblock


\bibitem[Wang et~al\mbox{.}(2025a)]%
        {Wang2025SimAIUA}
\bibfield{author}{\bibinfo{person}{Xizheng Wang}, \bibinfo{person}{Qingxu Li},
  \bibinfo{person}{Yichi Xu}, \bibinfo{person}{Gang Lu}, \bibinfo{person}{Dan
  Li}, \bibinfo{person}{Li Chen}, \bibinfo{person}{Heyang Zhou},
  \bibinfo{person}{Linkang Zheng}, \bibinfo{person}{Sen Zhang},
  \bibinfo{person}{Yikai Zhu}, \bibinfo{person}{Yang Liu},
  \bibinfo{person}{Pengcheng Zhang}, \bibinfo{person}{Kun Qian},
  \bibinfo{person}{Kunling He}, \bibinfo{person}{Jiaqi Gao},
  \bibinfo{person}{Ennan Zhai}, \bibinfo{person}{Dennis Cai}, {and}
  \bibinfo{person}{Binzhang Fu}.} \bibinfo{year}{2025}\natexlab{a}.
\newblock \showarticletitle{SimAI: Unifying Architecture Design and Performance
  Tuning for Large-Scale Large Language Model Training with Scalability and
  Precision}. In \bibinfo{booktitle}{\emph{Symposium on Networked Systems
  Design and Implementation}}.
\newblock
\urldef\tempurl%
\url{https://api.semanticscholar.org/CorpusID:278205358}
\showURL{%
\tempurl}


\bibitem[Wang et~al\mbox{.}(2025b)]%
        {wang2025omniwise}
\bibfield{author}{\bibinfo{person}{Zixian Wang}, \bibinfo{person}{Cole Ramos},
  \bibinfo{person}{Muhammad~A Awad}, {and} \bibinfo{person}{Keith Lowery}.}
  \bibinfo{year}{2025}\natexlab{b}.
\newblock \showarticletitle{Omniwise: Predicting GPU Kernels Performance with
  LLMs}.
\newblock \bibinfo{journal}{\emph{arXiv preprint arXiv:2506.20886}}
  (\bibinfo{year}{2025}).
\newblock


\bibitem[Williams et~al\mbox{.}(2009)]%
        {williams2009roofline}
\bibfield{author}{\bibinfo{person}{Samuel Williams}, \bibinfo{person}{Andrew
  Waterman}, {and} \bibinfo{person}{David Patterson}.}
  \bibinfo{year}{2009}\natexlab{}.
\newblock \showarticletitle{Roofline: an insightful visual performance model
  for multicore architectures}.
\newblock \bibinfo{journal}{\emph{Commun. ACM}} \bibinfo{volume}{52},
  \bibinfo{number}{4} (\bibinfo{year}{2009}), \bibinfo{pages}{65--76}.
\newblock


\bibitem[Wu et~al\mbox{.}(2025)]%
        {Wu2025TokenSimEH}
\bibfield{author}{\bibinfo{person}{Feiyang Wu}, \bibinfo{person}{Zhuohang
  Bian}, \bibinfo{person}{Guoyang Duan}, \bibinfo{person}{Tianle Xu},
  \bibinfo{person}{Junchi Wu}, \bibinfo{person}{Teng Ma},
  \bibinfo{person}{Yongqiang Yao}, \bibinfo{person}{Ruihao Gong}, {and}
  \bibinfo{person}{Youwei Zhuo}.} \bibinfo{year}{2025}\natexlab{}.
\newblock \showarticletitle{TokenSim: Enabling Hardware and Software
  Exploration for Large Language Model Inference Systems}. In
  \bibinfo{booktitle}{\emph{Advanced Parallel Programming Technologies}}.
\newblock
\urldef\tempurl%
\url{https://api.semanticscholar.org/CorpusID:276928157}
\showURL{%
\tempurl}


\bibitem[Wu et~al\mbox{.}(2022)]%
        {wu2022sparseloop}
\bibfield{author}{\bibinfo{person}{Yannan~Nellie Wu}, \bibinfo{person}{Po-An
  Tsai}, \bibinfo{person}{Angshuman Parashar}, \bibinfo{person}{Vivienne Sze},
  {and} \bibinfo{person}{Joel~S Emer}.} \bibinfo{year}{2022}\natexlab{}.
\newblock \showarticletitle{Sparseloop: An analytical approach to sparse tensor
  accelerator modeling}. In \bibinfo{booktitle}{\emph{2022 55th IEEE/ACM
  International Symposium on Microarchitecture (MICRO)}}. IEEE,
  \bibinfo{pages}{1377--1395}.
\newblock


\bibitem[Yarlagadda et~al\mbox{.}(2025)]%
        {yarlagadda2025maya}
\bibfield{author}{\bibinfo{person}{Srihas Yarlagadda}, \bibinfo{person}{Amey
  Agrawal}, \bibinfo{person}{Elton Pinto}, \bibinfo{person}{Hakesh Darapaneni},
  \bibinfo{person}{Mitali Meratwal}, \bibinfo{person}{Shivam Mittal},
  \bibinfo{person}{Pranavi Bajjuri}, \bibinfo{person}{Srinivas Sridharan},
  {and} \bibinfo{person}{Alexey Tumanov}.} \bibinfo{year}{2025}\natexlab{}.
\newblock \showarticletitle{Maya: Optimizing Deep Learning Training Workloads
  using GPU Runtime Emulation}.
\newblock \bibinfo{journal}{\emph{arXiv preprint arXiv:2503.20191}}
  (\bibinfo{year}{2025}).
\newblock


\bibitem[Yuan et~al\mbox{.}(2024)]%
        {yuan2024llm}
\bibfield{author}{\bibinfo{person}{Zhihang Yuan}, \bibinfo{person}{Yuzhang
  Shang}, \bibinfo{person}{Yang Zhou}, \bibinfo{person}{Zhen Dong},
  \bibinfo{person}{Zhe Zhou}, \bibinfo{person}{Chenhao Xue},
  \bibinfo{person}{Bingzhe Wu}, \bibinfo{person}{Zhikai Li},
  \bibinfo{person}{Qingyi Gu}, \bibinfo{person}{Yong~Jae Lee}, {et~al\mbox{.}}}
  \bibinfo{year}{2024}\natexlab{}.
\newblock \showarticletitle{Llm inference unveiled: Survey and roofline model
  insights}.
\newblock \bibinfo{journal}{\emph{arXiv preprint arXiv:2402.16363}}
  (\bibinfo{year}{2024}).
\newblock


\bibitem[Zeni et~al\mbox{.}(2024)]%
        {zeni2024starlight}
\bibfield{author}{\bibinfo{person}{Alberto Zeni}, \bibinfo{person}{Emanuele
  Del~Sozzo}, \bibinfo{person}{Eleonora D'Arnese}, \bibinfo{person}{Davide
  Conficconi}, {and} \bibinfo{person}{Marco~D Santambrogio}.}
  \bibinfo{year}{2024}\natexlab{}.
\newblock \showarticletitle{Starlight: A kernel optimizer for GPU processing}.
\newblock \bibinfo{journal}{\emph{J. Parallel and Distrib. Comput.}}
  \bibinfo{volume}{187} (\bibinfo{year}{2024}), \bibinfo{pages}{104832}.
\newblock


\bibitem[Zhang et~al\mbox{.}(2026)]%
        {pipeweave2026}
\bibfield{author}{\bibinfo{person}{Kaixuan Zhang}, \bibinfo{person}{Yunfan
  Cui}, \bibinfo{person}{Shuhao Zhang}, \bibinfo{person}{Chutong Ding},
  \bibinfo{person}{Shiyou Qian}, \bibinfo{person}{Luping Wang},
  \bibinfo{person}{Jian Cao}, \bibinfo{person}{Guangtao Xue},
  \bibinfo{person}{Cheng Huang}, \bibinfo{person}{Guodong Yang}, {and}
  \bibinfo{person}{Liping Zhang}.} \bibinfo{year}{2026}\natexlab{}.
\newblock \showarticletitle{PipeWeave: Synergizing Analytical and Learning
  Models for Unified GPU Performance Prediction}.
\newblock \bibinfo{journal}{\emph{arXiv preprint}} (\bibinfo{year}{2026}).
\newblock
\urldef\tempurl%
\url{https://arxiv.org/abs/2601.14910}
\showURL{%
\tempurl}


\bibitem[Zheng et~al\mbox{.}(2023)]%
        {zheng2023tileflow}
\bibfield{author}{\bibinfo{person}{Size Zheng}, \bibinfo{person}{Siyuan Chen},
  \bibinfo{person}{Siyuan Gao}, \bibinfo{person}{Liancheng Jia},
  \bibinfo{person}{Guangyu Sun}, \bibinfo{person}{Runsheng Wang}, {and}
  \bibinfo{person}{Yun Liang}.} \bibinfo{year}{2023}\natexlab{}.
\newblock \showarticletitle{TileFlow: A Framework for Modeling Fusion Dataflow
  via Tree-based Analysis}. In \bibinfo{booktitle}{\emph{Proceedings of the
  56th Annual IEEE/ACM International Symposium on Microarchitecture}}.
  \bibinfo{pages}{1271--1288}.
\newblock


\bibitem[Zheng et~al\mbox{.}(2025)]%
        {zheng2025tilelink}
\bibfield{author}{\bibinfo{person}{Size Zheng}, \bibinfo{person}{Jin Fang},
  \bibinfo{person}{Xuegui Zheng}, \bibinfo{person}{Qi Hou},
  \bibinfo{person}{Wenlei Bao}, \bibinfo{person}{Ningxin Zheng},
  \bibinfo{person}{Ziheng Jiang}, \bibinfo{person}{Dongyang Wang},
  \bibinfo{person}{Jianxi Ye}, \bibinfo{person}{Haibin Lin}, {et~al\mbox{.}}}
  \bibinfo{year}{2025}\natexlab{}.
\newblock \showarticletitle{Tilelink: Generating efficient
  compute-communication overlapping kernels using tile-centric primitives}.
\newblock \bibinfo{journal}{\emph{arXiv preprint arXiv:2503.20313}}
  (\bibinfo{year}{2025}).
\newblock


\bibitem[Zhong et~al\mbox{.}(2024)]%
        {zhong2024distserve}
\bibfield{author}{\bibinfo{person}{Yinmin Zhong}, \bibinfo{person}{Shengyu
  Liu}, \bibinfo{person}{Junda Chen}, \bibinfo{person}{Jianbo Hu},
  \bibinfo{person}{Yibo Zhu}, \bibinfo{person}{Xuanzhe Liu},
  \bibinfo{person}{Xin Jin}, {and} \bibinfo{person}{Hao Zhang}.}
  \bibinfo{year}{2024}\natexlab{}.
\newblock \showarticletitle{$\{$DistServe$\}$: Disaggregating prefill and
  decoding for goodput-optimized large language model serving}. In
  \bibinfo{booktitle}{\emph{18th USENIX Symposium on Operating Systems Design
  and Implementation (OSDI 24)}}. \bibinfo{pages}{193--210}.
\newblock


\bibitem[Zhou et~al\mbox{.}(2021b)]%
        {zhou2021gpa}
\bibfield{author}{\bibinfo{person}{Keren Zhou}, \bibinfo{person}{Xiaozhu Meng},
  \bibinfo{person}{Ryuichi Sai}, \bibinfo{person}{Dejan Grubisic}, {and}
  \bibinfo{person}{John Mellor-Crummey}.} \bibinfo{year}{2021}\natexlab{b}.
\newblock \showarticletitle{An automated tool for analysis and tuning of
  gpu-accelerated code in hpc applications}.
\newblock \bibinfo{journal}{\emph{IEEE Transactions on Parallel and Distributed
  Systems}} \bibinfo{volume}{33}, \bibinfo{number}{4} (\bibinfo{year}{2021}),
  \bibinfo{pages}{854--865}.
\newblock


\bibitem[Zhou et~al\mbox{.}(2021a)]%
        {zhou2020gpa_cgo}
\bibfield{author}{\bibinfo{person}{Keren Zhou}, \bibinfo{person}{Xiaozhu Meng},
  \bibinfo{person}{Ryuichi Sai}, {and} \bibinfo{person}{John Mellor-Crummey}.}
  \bibinfo{year}{2021}\natexlab{a}.
\newblock \showarticletitle{GPA: A GPU Performance Advisor Based on Instruction
  Sampling}. In \bibinfo{booktitle}{\emph{2021 IEEE/ACM International Symposium
  on Code Generation and Optimization (CGO)}}. \bibinfo{pages}{115--125}.
\newblock


\end{thebibliography}

\end{document}